\documentclass[aps,prb,10pt,twocolumn,superscriptaddress]{revtex4-1}

\usepackage{graphicx}
\usepackage{amssymb,amsmath,mathtools,textcomp}
\usepackage{bbm,enumerate}
\usepackage[colorlinks,urlcolor=blue,linkcolor=blue,anchorcolor=blue,citecolor=blue]{hyperref}

\newcommand{\nio}{Na$_2$IrO$_3$}
\newcommand{\lio}{Li$_2$IrO$_3$}
\newcommand{\aio}{$A_2$IrO$_3$}
\newcommand{\rucl}{$\alpha$-RuCl$_3$}

\newcommand{\jeff}{j_{\rm eff}}

\newcommand{\hk}{Heisenberg-Kitaev}

\newcommand{\hc}{h_{0}}

\newcommand{\llangle}{\langle\!\langle}
\newcommand{\rrangle}{\rangle\!\rangle}

\newcommand{\lllangle}{\langle\!\langle\!\langle}
\newcommand{\rrrangle}{\rangle\!\rangle\!\rangle}

\graphicspath{{./}{./plots/}}

\begin{document}

\title{
Magnetization processes of zigzag states on the honeycomb lattice: \texorpdfstring{\\}{}
Identifying spin models for \texorpdfstring{\rucl}{RuCl3} and \texorpdfstring{\nio}{Na2IrO3}
}

\author{Lukas Janssen}
\affiliation{Institut f\"ur Theoretische Physik, Technische Universit\"at Dresden,
01062 Dresden, Germany}

\author{Eric C.\ Andrade}
\affiliation{Instituto de F\'{i}sica de S\~ao Carlos, Universidade de S\~ao Paulo, C.P. 369,
S\~ao Carlos, SP,  13560-970, Brazil}

\author{Matthias Vojta}
\affiliation{Institut f\"ur Theoretische Physik, Technische Universit\"at Dresden,
01062 Dresden, Germany}

\begin{abstract}
We study the field-induced magnetization processes of extended Heisenberg-Kitaev models on the honeycomb lattice, taking into account off-diagonal and longer-range exchange interactions, using a combination of Monte-Carlo simulations, classical energy minimization, and spin-wave theory.
We consider a number of different parameter sets, previously proposed to describe the magnetic behavior of $\alpha$-RuCl$_3$ and Na$_2$IrO$_3$ with their antiferromagnetic zigzag ground states.
By classifying these parameter sets, we reveal the existence of three distinct mechanisms to stabilize zigzag states, which differ in the sign of the nearest-neighbor Kitaev interaction, the role of longer-range interactions, and the magnitude of the off-diagonal $\Gamma_1$ interaction. While experimentally hardly distinguishable at zero field, we find that the three different scenarios lead to significantly different magnetization processes in applied magnetic fields.
In particular, we show that a sizable off-diagonal interaction $\Gamma_1 > 0$ naturally explains the strongly anisotropic field responses observed in $\alpha$-RuCl$_3$ without the need for a strong anisotropy in the effective $g$ tensor. Moreover, for a generic field direction, it leads to a high-field state with a finite transversal magnetization, which should be observable in $\alpha$-RuCl$_3$.
\end{abstract}

\maketitle

%%%%%%%%%%%%%%%%%%%%%%%%%%%%%%%%%%%%%%%%%%%%%%%%%%%%%%%%%%%%%%%%%%%%
%%%%%%%%%%%%%%%%%%%%%%%%%%%%%%%%%%%%%%%%%%%%%%%%%%%%%%%%%%%%%%%%%%%%
%%%%%%%%%%%%%%%%%%%%%%%%%%%%%%%%%%%%%%%%%%%%%%%%%%%%%%%%%%%%%%%%%%%%

\section{Introduction}

Strongly-correlated electron systems in the regime of strong spin-orbit coupling have recently attracted a lot of attention, luring with the prospect of material realizations of nontrivial phases with possibly exotic excitations. Candidate systems that are presently intensely studied, both experimentally and theoretically, are based on transition metal ions with partially filled $4d$ or $5d$ shells, such as \rucl\ and {\aio} ($A = \mathrm{Na}, \mathrm{Li}$).\cite{trebst2017} In these systems, the transition-metal ions are caged in chlorine RuCl$_6$ and oxygen IrO$_6$ octahedra, respectively. These cages share an edge with each of their three neighboring octahedra, such that the central ions form weakly coupled layers of honeycomb lattices. The combined effect of level splitting due to the octahedral crystal field and spin-orbit coupling generates $\jeff = 1/2$ moments. Additional Coulomb repulsion drives the system into a Mott insulating state already above room temperature, suggesting a local-moment description as a low-energy model.\cite{plumb2014, choi2012} In the simplest, nondistorted cubic geometry, the edge-sharing octrahedra allow two 90\textdegree\ Ru-Cl-Ru and Ir-O-Ir exchange paths, respectively. This leads to a partial destructive interference of the symmetric Heisenberg interaction in favor of a dominant bond-directional Ising-type exchange of the form of the celebrated Kitaev compass model,\cite{jackeli2009, chaloupka2010} fueling the hope to realize Kitaev's spin liquid phase of Majorana fermions.\cite{kitaev2006}

Experimentally, {\rucl} and {\nio} show an ordering transition towards an antiferromagnetic state of collinear ``zigzag'' type,\cite{choi2012, sears2015, johnson2015} while $\alpha$-{\lio} exhibits an incommensurate spiral magnetic ground state. \cite{williams2016, footnote1} In all cases, the N\'eel temperature is small compared to the Curie-Weiss temperature, indicating substantial frustration.
A suitable perturbation that allows to stabilize the spin-liquid state experimentally is therefore much sought-after.\cite{yadav2016} Interestingly, in {\rucl} a number of very recent measurements indeed appear to be consistent with a, possibly continuous,\cite{wolter2017} transition from the ordered phase towards a quantum paramagnetic phase, driven by magnetic field\cite{leahy2017, baek2017, sears2017, zheng2017, hentrich2017, banerjee2017} or external pressure.\cite{wang2017, cui2017}

At zero field and ambient pressure, magnetic order must be caused by interactions beyond the nearest-neighbor Kitaev model,\cite{chaloupka2010} and it has been pointed out that longer-range\cite{singh2012} as well as off-diagonal interactions\cite{rau2014} may play a role. An additional complication is that the materials appear to realize a low-temperature monoclinic $C2/m$ crystal structure instead of the idealized cubic one,\cite{choi2012, johnson2015} however, the actual size and role of the trigonal distortion is disputed.\cite{yadav2016, park2016, tjeng2017}
Furthermore, when subject to an external magnetic field, \rucl, for instance, unveils a huge anisotropy between in-plane and out-of-plane magnetic responses. \cite{johnson2015, majumder2015, kubota2015} In the spin models with only Kitaev and Heisenberg interactions, such large anisotropy can only arise from largely anisotropic $g$ factors.\cite{kubota2015} This, however, would require a substantial trigonal distortion, not easily consistent with crystallographic measurements.\cite{cao2016, park2016, tjeng2017} Evidently, a more natural explanation for the magnetic anisotropy would be an effective spin model which displays a strong \emph{intrinsic} magnetic anisotropy, present even if the $g$ tensor were entirely isotropic. To the best of our knowledge, such has so far not been demonstrated.

To date, the debate about the most appropriate spin model for {\rucl} and {\nio} is by no means settled. There are basically (at least) three scenarios that were suggested to stabilize the experimentally observed zigzag order in these systems:
\begin{description}
\item[Scenario 1] If the most dominant interactions are between nearest neighbors only, a zigzag-ordered state occurs if the Kitaev interaction is \emph{antiferromagnetic}, $K_1\!>\!0$.\cite{chaloupka2013}
\item[Scenario 2] Zigzag order can be stabilized also for \emph{ferromagnetic} Kitaev coupling, $K_1\!<\!0$, if a sizable third-neighbor antiferromagnetic Heisenberg interaction, $J_3\!>\!0$, is present.\cite{fouet2001, winter2016}
\item[Scenario 3] Zigzag order has furthermore been suggested to be realizable also in a nearest-neighbor model with \emph{ferromagnetic} Kitaev coupling, $K_1\!<\!0$, when supplemented with a sizable off-diagonal interaction $\Gamma_1\!>\!0$.\cite{rau2014, ran2017, wang2016}
\end{description}
For \rucl, for instance, inelastic neutron-scattering data have been argued to be consistent with antiferromagnetic Kitaev coupling $K_1$ (Scenario 1).\cite{banerjee2016a, banerjee2016b, gohlke2017}
In contrast, \textit{ab-initio} studies typically find a ferromagnetic $K_1$ and have emphasized the significance of longer-range couplings (Scenario 2).\cite{winter2016, kim2016, yadav2016}
Finally, some recent works, based on \textit{ab-initio} calculations or fits to neutron-scattering data, suggest by contrast a nearest-neighbor $K_1$-$\Gamma_1$ model with ferromagnetic $K_1$ and strong positive $\Gamma_1$ as a minimal description of \rucl\ (Scenario 3).\cite{ran2017, wang2016} (Notably, we will demonstrate below that the plain $K_1$-$\Gamma_1$ model is \textit{insufficient} to stabilize zigzag order and needs to be supplemented with, e.g., a finite ferromagnetic Heisenberg coupling $J_1 < 0$.) A similar $K_1$-$\Gamma_1$ model with $K_1 < 0$ and strong $\Gamma_1 > 0$, in this case supplemented with small Heisenberg interactions $J_1 < 0$ and $J_3 > 0$ (stabilizing zigzag order) has recently also been proposed.\cite{winter2017}
Experimentally, it is difficult to differentiate between these scenarios at zero field. A possible indicator is the moment direction in the ordered state;\cite{sizyuk2016, chaloupka2016} however, this observable is difficult to access and requires the use of polarized neutrons\cite{cao2016} or magnetic x-ray scattering data.\cite{chun2015} Clearly, further theoretical predictions that distinguish between these different mechanisms are necessary to resolve the debate.

In this paper, we demonstrate that the response of the system to an external field differs substantially for the different scenarios of stabilizing the zigzag state, and we argue that this can be used to narrow down the experimentally relevant parameter range of the models.
Recently, we have mapped out the magnetization processes in the nearest-neighbor Heisenberg-Kitaev model.\cite{janssen2016} Here, we contrast these findings to new results that are obtained within the extended Heisenberg-Kitaev models including longer-range and off-diagonal interactions. Given the huge parameter space, we mainly restrict our attention to those parameter regimes which lead to zigzag order in zero field, but some of the analytical considerations are more general, as will be noted below. Our methodology is concerned with the semiclassical limit of large spin $S$, but we also compute quantum corrections to the high-field magnetization, which we evaluate for $S=1/2$.

We devote a specific discussion to the effect of large $\Gamma_1$ in the presence of a magnetic field. We argue that this naturally leads to a strongly anisotropic field response, and we also discuss under which circumstances the $\Gamma_1$ term induces a finite transversal magnetization even in the high-field phase.

The remainder of the paper is organized as follows:
In Sec.~\ref{sec:model}, we introduce the general honeycomb-lattice spin Hamiltonian together with the different parameter sets considered in this paper.
Sec.~\ref{sec:methods} quickly summarizes the different methods used to tackle the problem.
Secs.~\ref{sec:zigzag1}, \ref{sec:zigzag2}, and \ref{sec:zigzag3} discuss the influence of a magnetic field on the zigzag states stabilized by the three different mechanisms, corresponding to different parameter regimes of the general model. In all cases, we discuss degeneracies and the expected low-field behavior using analytical arguments. We also show extensive numerical results for the magnetization processes for various concrete parameter sets.
Sec.~\ref{sec:gamma} contains a specific discussion of the effects of a large off-diagonal $\Gamma_1$ interaction.
A summary of our results, together with a discussion of their experimental relevance and suggestions for future experiments, closes the main part of the paper.
In the appendix, we demonstrate that a nearest-neighbor model with ferromagnetic $K_1 < 0$ and strong $\Gamma_1 > 0$ stabilizes zigzag order only when supplemented with a finite $J_1 < 0$.

%%%%%%%%%%%%%%%%%%%%%%%%%%%%%%%%%%%%%%%%%%%%%%%%%%%%%%%%%%%%%%%%%%%%
%%%%%%%%%%%%%%%%%%%%%%%%%%%%%%%%%%%%%%%%%%%%%%%%%%%%%%%%%%%%%%%%%%%%
%%%%%%%%%%%%%%%%%%%%%%%%%%%%%%%%%%%%%%%%%%%%%%%%%%%%%%%%%%%%%%%%%%%%

\section{Models}
\label{sec:model}

\subsection{Generic Hamiltonian}

Out of the large variety of spin Hamiltonians proposed for {\rucl} and {\nio}, most can be written as extensions of the {\hk} model originally proposed in Ref.~\onlinecite{chaloupka2010}. Consequently, we consider models on the honeycomb lattice of the general form
\begin{align} \label{eq:hamiltonian-1}
\mathcal{H}_0 & =
\sum_{\langle ij\rangle_\gamma} \left[
J_1 \vec S_i \cdot \vec S_j  + K_1 S_i^\gamma S_j^\gamma
+ \Gamma_1 \left( S_i^\alpha S_j^\beta + S_i^\beta S_j^\alpha\right)
\right]
\nonumber \\
&+ \sum_{\llangle ij \rrangle_\gamma} \left(
J_2 \vec S_i \cdot \vec S_j  + K_2 S_i^\gamma S_j^\gamma
\right)
%
% \nonumber \\ & \quad
%
+ \sum_{\lllangle ij\rrrangle}
J_3 \vec S_i \cdot \vec S_j.
\end{align}
Here, $J_{1,2,3}$ correspond to first-, second-, and third-neighbor Heisenberg couplings, similarly $K_{1,2}$ are Kitaev couplings, and $\Gamma_{1}$ is a symmetric off-diagonal coupling. $\langle i j \rangle_\gamma$ and $\llangle ij \rrangle_\gamma$ denote first- and second-neighbor $\gamma$-bonds, respectively, with $\gamma = x,y,z$. On $z$ bonds $(\alpha,\beta,\gamma) = (x,y,z)$, with cyclic permutation for $x$ and $y$ bonds. The third-neighbor bonds $\lllangle ij \rrrangle$ are along opposite points of the same hexagon. We assume cubic symmetry, corresponding to a perfect honeycomb structure in $\mathcal{H}_0$, and neglect possible trigonal distortions.

\begin{figure*}
 \includegraphics[width=\linewidth]{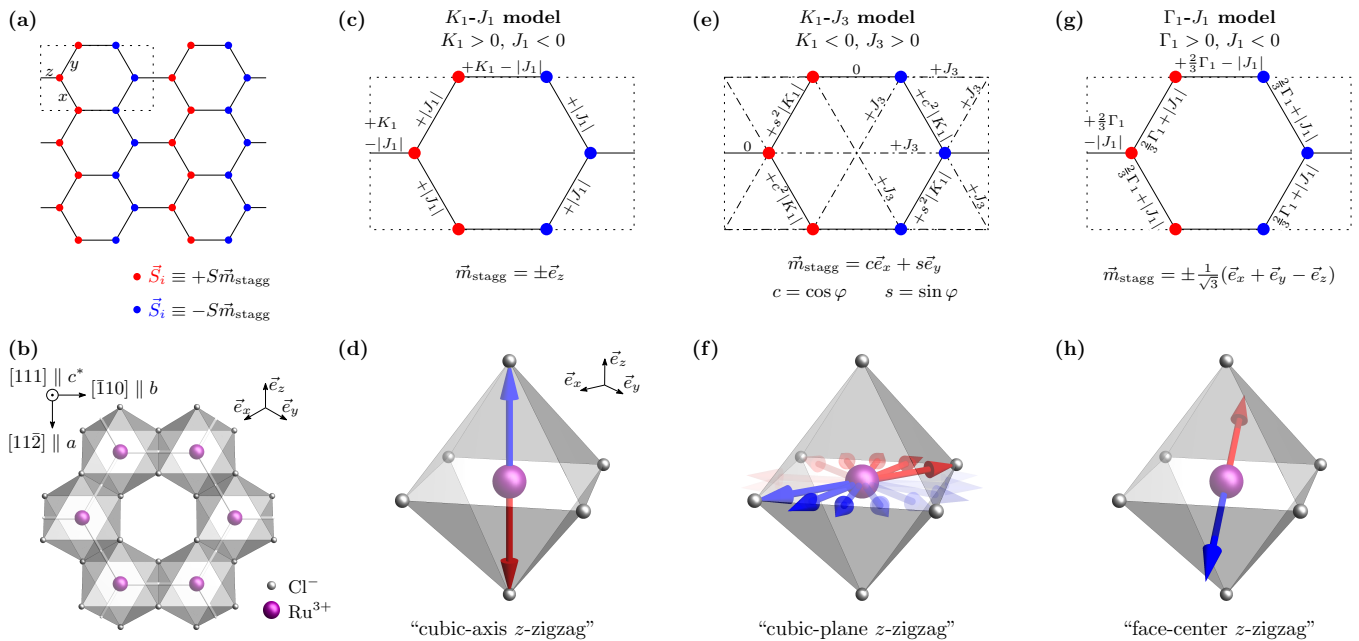}
 \caption{{(a)} Ferromagnetic zigzag chains along $x$ and $y$ bonds with antiferromagnetically ordered $z$ bonds (``$z$-zigzag''), together with the definition of the staggered magnetization $\vec m_\text{stagg}$. Dotted rectangle: magnetic unit cell.
{(b)} Hexagonal structure of Ru$^{3+}$ ions with edge-shared octahedra of Cl$^-$ ions. The cubic axes $\vec e_x$, $\vec e_y$, and $\vec e_z$ are along Ru-Cl bonds, while the $[\bar 110]$ direction is along Ru-Ru $z$-bonds of the honeycomb plane. In the conventions of Ref.~\onlinecite{cao2016} the latter corresponds to the crystallographic $b$ axis, while the $[11\bar 2]$ in our cubic basis then coincides with the $a$ axis, with $a \perp b$. The $[111]$ direction is perpendicular to the honeycomb plane and is sometimes referred to as $c^*$ axis.\cite{johnson2015}
 {(c)} Classical energy gain on each bond in Scenario~1 (antiferromagnetic $K_1$, ferromagnetic $J_1$).
 In the $z$-zigzag state, $\vec m_\text{stagg}$ is parallel or antiparallel to the $z$ axis  (``cubic-axis $z$-zigzag''). In \rucl, the spins point along Ru-Cl bonds  {(d)}.
 {(e)} Same as (c), but in Scenario~2 (ferromagnetic $K_1$, antiferromagnetic $J_3$). In the $z$-zigzag state, $\vec m_\text{stagg}$ can (classically) point everywhere in the $xy$ plane (``cubic-plane $z$-zigzag'').  In \rucl, this corresponds to the planes perpendicular to the Ru-Cl bonds {(f)}.
 {(g)} Same as (e), but in Scenario~3 (strong $\Gamma_1>0$). Here, we display the limit $\Gamma_1 \gg |K_1|$ with $\Gamma_1/ (-J_1) > 0$ finite. In the $z$-zigzag state, $\vec m_\text{stagg}$ lies in the $[11\bar1]$ direction. In \rucl, the spins point towards the centers of opposite faces of Cl$^-$ octahedra surrounding each Ru$^{3+}$ ion (``face-center $z$-zigzag'')   (h).
}
\label{fig:zigzag}
\end{figure*}

We are interested in the effects of a uniform external magnetic field, which couples linearly to the spins via
\begin{align} \label{eq:hamiltonian-2}
 \mathcal{H}_{\vec h} = - \vec h \cdot \sum_{i} \vec S_i,
\end{align}
where the constants have been absorbed into the field $\vec h \coloneqq g \mu_\mathrm{B} \mu_0 \vec H$, with $g$ the (possibly anisotropic) effective $g$ tensor and $\mu_\mathrm{B}$ the Bohr magneton.
The total Hamiltonian then is $\mathcal H = \mathcal H_0 + \mathcal H_{\vec h}$.

\subsection{Conventions for field directions}

Due to the broken spin rotation symmetry, and as will be demonstrated explicitly below, the response of the system to an external field $\vec h$ crucially depends on the field direction $\vec h/h$. For convenience, we will give all field directions in the cubic spin basis $\{\vec e_x, \vec e_y, \vec e_z\}$ and label them in the form $[xyz]$, i.e.,
\begin{equation}
\vec h \parallel [xyz] \quad \Leftrightarrow \quad \vec h \propto x \vec e_x + y \vec e_y + z \vec e_z.
\end{equation}
In RuCl$_3$, the cubic axes $\vec e_x$, $\vec e_y$, and $\vec e_z$ point along nearest-neighbor Ru-Cl bonds, see Fig.~\ref{fig:zigzag}(b). Consequently, the two in-plane field directions along the crystallographic $a$ and $b$ axes correspond to $\vec h \parallel [11\bar2]$  and $\vec h \parallel [\bar 110]$, respectively. The direction perpendicular to the honeycomb layer, which is sometimes referred to as $c^*$ axis,\cite{johnson2015} is labelled as the $[111]$ direction in our conventions. In the same way, $\vec h \parallel [001]$ denotes an intermediate direction which lies in the $ac$ plane and is tilted 55\textdegree\ away from the $c^*$ axis towards the $-a$ axis.

\subsection{Parameter sets}

\begin{table*}
\caption{Parameter sets for spin-spin exchange interactions in {\rucl} and {\nio} from the literature.
Refs.~\onlinecite{banerjee2016a, banerjee2016b, ran2017, winter2017} apply fits to neutron-scattering data within different minimal-model descriptions.
Ref.~\onlinecite{chaloupka2013} uses the magnetic susceptibility curve as an additional information.
Refs.~\onlinecite{kim2015, kim2016, winter2016, wang2016} derive hopping integrals from density functional theory (DFT) and utilize them to compute magnetic interactions by exact-diagonalization techniques and strong-coupling $t/U$ expansion, respectively.
Refs.~\onlinecite{sizyuk2014, sizyuk2016} use hopping parameters from the DFT calculations of Ref.~\onlinecite{foyevtsova2013} and apply constraints from the experimentally observed direction of the ordered moments in the zigzag state.
Refs.~\onlinecite{katukuri2014, yadav2016} exploit the multireference configuration interaction (MRCI) quantum chemistry technique to compute the nearest-neighbor interactions,\cite{footnote3} supplemented by a fit to the Curie-Weiss temperature $\theta_\text{CW}$ and the magnetization curve, respectively, to estimate $J_2$ and $J_3$.
We note that Refs.~\onlinecite{yadav2016, kim2016} suggest non-negligible anisotropies in some magnetic interactions which are not explicitly displayed in the table.
``---''~denotes magnetic interactions that were not computed within the respective approach.
}
\label{tab:par}
 \begin{tabular*}{\textwidth}{@{\extracolsep{\fill} }lcccccccp{13em}cc} %16em
  \hline\hline
  Set & Material & $J_1$ [meV] & $K_1$ [meV] & $\Gamma_1$ [meV] &
  $J_2$ [meV] & $K_2$ [meV] & $J_3$ [meV] & \multicolumn{1}{c}{Method} & Ref. & Year
  \\ \hline
  1 & \rucl & $-4.6$ & $+7.0$ & --- & --- & --- & --- & fit to neutron scattering &\onlinecite{banerjee2016a}, \onlinecite{banerjee2016b} & 2016 \\
  1' & \nio & $-4.0$ & $+10.5$ & --- & --- & --- & --- & fit to susceptibility \& neutron scattering &
  \onlinecite{chaloupka2013} & 2013 \\
  1+$\Gamma$ & \rucl & $-12$ & $+17$ & $+12$ & --- & --- & --- & DFT + $t/U$ expansion &\onlinecite{kim2015} & 2015 \\
  2 & \nio & $0$ & $-17$ & $0$ & $0$ & --- & $+6.8$ & DFT + exact diagonalization &
  \onlinecite{winter2016} & 2016 \\
  2+$\Gamma$ & \nio & $+3$ & $-17$ & $+1$ & $-3$ & $+6$ & $+1$ & DFT + $t/U$ expansion, direction of moments &
  \onlinecite{sizyuk2016}, \onlinecite{sizyuk2014} & 2016 \\
  (2+$\Gamma$)' & \nio & $+3$ & $-17.5$ & $+1$ & $+5$ & --- & $+5$ & MRCI, fit to $\theta_\text{CW}$&
  \onlinecite{katukuri2014} & 2014 \\
  (2+$\Gamma$)'' & \rucl & $+1.2$ & $-5.6$ & $+1$ & $+0.3$ & --- & $+0.3$ & MRCI, fit to magnetization &
  \onlinecite{yadav2016} & 2016 \\
  2/3 & \rucl & $-1.7$ & $-6.6$ & $+6.6$ & $0$ & --- & $+2.7$ & DFT + exact diagonalization &
  \onlinecite{winter2016} & 2016 \\
  3 & \rucl & --- & $-6.8$ & $+9.5$ & --- & --- & --- & fit to neutron scattering &
  \onlinecite{ran2017} & 2017 \\
  3' & \rucl & --- & $-5.5$ & $+7.6$ & --- & --- & --- & DFT + $t/U$ expansion &
  \onlinecite{wang2016} & 2016 \\
  3'' & \rucl & $-1$ & $-8$ & $+4$ & --- & --- & --- & DFT + $t/U$ expansion &
  \onlinecite{kim2016} & 2016 \\
  3+$J_3$ & \rucl & $-0.5$ & $-5.0$ & $+2.5$ & --- & --- & $+0.5$ & fit to neutron scattering & \onlinecite{winter2017} & 2017 \\
  \hline\hline
 \end{tabular*}
\end{table*}

In Table \ref{tab:par}, we list popular parameter sets for the couplings $J_{1,2,3}$, $K_{1,2}$, and $\Gamma_1$, that were suggested either on the basis of \textit{ab-initio} calculations or by fitting the predictions of different simplified versions of the above model to experimental data.
The parameter sets can roughly be divided into three groups, corresponding to the three scenarios listed in the introduction:
\begin{enumerate}[(1)]
\item Dominant antiferromagnetic $K_1>0$, supplemented by ferromagnetic $J_1<0$ and small longer-ranged interactions,
\item Dominant ferromagnetic $K_1<0$ together with large antiferromagnetic $J_3>0$,
\item Strong $\Gamma_1>0$ in conjunction with ferromagnetic~$K_1<0$ and only small Heisenberg couplings.
\end{enumerate}
The classical energy contributions to a zero-field zigzag state within three different minimal models, representative for the three scenarios, are illustrated in Fig.~\ref{fig:zigzag}. The observed zigzag states in \rucl\ (Refs.~\onlinecite{sears2015, johnson2015}) and \nio\ (Ref.~\onlinecite{choi2012}) are in principle compatible with all three groups.
We reiterate, though, that the plain $K_1$-$\Gamma_1$ model suggested in Refs.~\onlinecite{ran2017, wang2016} needs to be supplemented with, e.g., a finite ferromagnetic Heisenberg coupling $J_1 < 0$ in order to stabilize zigzag order, see Sec.~\ref{sec:zigzag3} and the appendix for details. Parameter Sets 3 and 3' therefore do \emph{not} lead to a zigzag ground state, at least on the classical level.

In the absence of an external magnetic field, it appears difficult to experimentally rule out any of the various (mutually incompatible) parameter sets. In what follows, we argue that the behavior in an external field qualitatively differs for the three scenarios. In particular, we propose the magnetic field response as an experimental indicator of the sign of $K_1$ and demonstrate that the value of $\Gamma_1$ can be obtained (i) by measuring the angle dependence of the magnetic susceptibility or critical field strength, or (ii) by applying a strong magnetic field provided that the finite transversal magnetization is accessible in the experimental setup.

%%%%%%%%%%%%%%%%%%%%%%%%%%%%%%%%%%%%%%%%%%%%%%%%%%%%%%%%%%%%%%%%%%%%
%%%%%%%%%%%%%%%%%%%%%%%%%%%%%%%%%%%%%%%%%%%%%%%%%%%%%%%%%%%%%%%%%%%%
%%%%%%%%%%%%%%%%%%%%%%%%%%%%%%%%%%%%%%%%%%%%%%%%%%%%%%%%%%%%%%%%%%%%

\section{Methods}
\label{sec:methods}

\subsection{Monte-Carlo simulations}

To obtain concrete results for the magnetization processes of different microscopic models, we have employed classical Monte-Carlo (MC) simulations, combining single-site and parallel-tempering updates in order to equilibrate the spin configurations at low $T$; for details see Ref.~\onlinecite{janssen2016}.
Here, we have performed field scans on lattices with up to $N = 2 \times 18^2$ sites at temperatures down to $T/ |K_1 S^2| \simeq 10^{-3}$ and have computed the static spin structure factor $S_{\vec k} = N^{-1} \sum_{ij}\langle \vec S_i \cdot \vec S_j \rangle e^{i \vec k \cdot (\vec R_i - \vec R_j)}$, where $\vec R_i$ is the lattice vector at site $i$. The lowest-$T$ configurations have been further cooled down to obtain the zero-temperature magnetization $\vec m$ in the respective classical ground state.

\subsection{Variational approach}

The MC simulations are augmented with a semi-analytical variational calculation in which we have assumed a four-site magnetic unit cell with two pairs of pairwise parallel spins according to the three proposed \mbox{$x$-,} \mbox{$y$-,} and $z$-zigzag patterns (Fig.~\ref{fig:zigzag}). The classical energy has been minimized with respect to the four independent spherical angles $\vartheta_{1,2}$ and $\varphi_{1,2}$ of the two pairs of spins for each zigzag pattern. We have cross-checked the variational results with our MC data points. For all cases studied the ground state is indeed either a (uniformly or non-uniformly) canted zigzag state or a homogeneous high-field state, and we find perfect agreement within a relative uncertainty of typically $\Delta |\vec m \cdot \hat h|/|\vec m \cdot \hat h| \lesssim 10^{-7}$ (except for a few data points that are very close to first-order transitions). This way, we obtain the full classical magnetization curve $\vec m$ as function of continuous $h$ for various field directions including $[11\bar 2]$ (corresponding to the $a$ axis in \rucl), $[\bar 110]$ ($b$ axis), $[111]$ ($c^*$ axis), and $[001]$ (intermediate direction in the $ac$ plane).

\subsection{Spin-wave theory}

As another cross-check, and in order to obtain an estimate on the influence of quantum fluctuations in the respective $S=1/2$ systems, we employ spin-wave theory in the homogeneous high-field phase. For large $h > \hc$, the magnon spectrum is gapped. Upon lowering the field strength, the gap decreases. If the transition towards the canted zigzag phase at $\hc$ is continuous, it can be understood as condensation of magnons, and the critical field strength at which the gap closes can be computed analytically.\cite{janssen2016} In linear spin-wave theory, it precisely agrees with the classical continuous transition point, and we have cross-checked our spin-wave results against the MC simulations and the variational technique.
In the cases of continuous transitions, we have checked that the instability wave vector, at which the magnon gap closes, agrees with the ordering wave vector of the canted zigzag state, as it should be as long as there are no intermediate phases with other ordering wave vectors.\cite{janssen2016}
Furthermore, we have computed the leading-order correction to the classical magnetization in the high-field phase for $\vec h \in ab$ and for $\vec h \perp ab$, the sizes of which allow us to assess the validity of our classical approximation.
Details on the spin-wave calculations are given in the Supplemental Materials to Refs.~\onlinecite{janssen2016, wolter2017}.

%%%%%%%%%%%%%%%%%%%%%%%%%%%%%%%%%%%%%%%%%%%%%%%%%%%%%%%%%%%%%%%%%%%%%%%%%%%
%%%%%%%%%%%%%%%%%%%%%%%%%%%%%%%%%%%%%%%%%%%%%%%%%%%%%%%%%%%%%%%%%%%%%%%%%%%
%%%%%%%%%%%%%%%%%%%%%%%%%%%%%%%%%%%%%%%%%%%%%%%%%%%%%%%%%%%%%%%%%%%%%%%%%%%

\section{Scenario 1: Zigzag order from antiferromagnetic \texorpdfstring{$K_1$}{K1} and ferromagnetic \texorpdfstring{$J_1$}{J1}}
\label{sec:zigzag1}

\begin{figure*}[t]
 \includegraphics[scale=0.75]{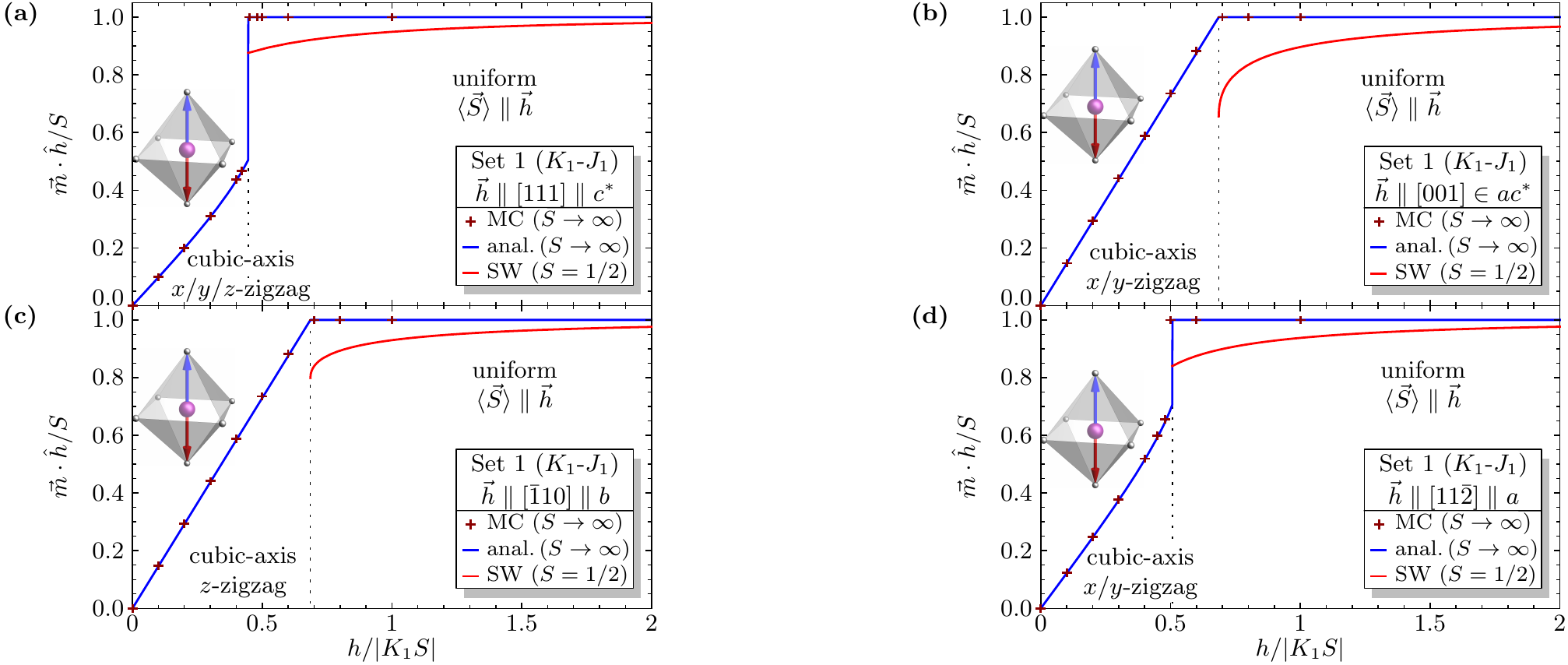}
\caption{
Magnetization in field direction for $J_1$-$K_1$ model with large $K_1 > 0$ and small $J_1 < 0$, using Parameter Set~1, from MC simulations (red crosses) and semi-analytical variational calculation (blue/dark curves). For comparison: spin-wave magnetization for $S=1/2$ (red/light curves, high-field phases only).
The small octahedra are given as a reminder of the spin alignment at zero field, cf.\ Fig.~\ref{fig:zigzag}.
For $\vec h \parallel [001]$ (b) and $\vec h \parallel [\bar 110]$ (c), the classical magnetization curves are linear in the intermediate-field regime and exhibit a continuous transition towards the uniform high-field phase at $\hc$, indicating a uniform canting mechanism with low-field spin alignment $\vec S_i \perp \vec h$ for $h \searrow 0$. For $\vec h \parallel [111]$ (a) and $\vec h \parallel [11\bar 2]$ (d), the magnetization curve becomes nonlinear and a direct continuous transition towards the high-field phase becomes impossible.
}
\label{fig:mag_set1}
\end{figure*}

In this and the following two sections, we discuss and contrast the different stabilization mechanisms of zigzag order corresponding to the three scenarios, and map out the consequential different magnetization processes.
We remind the reader that zigzag order is characterized by ferromagnetic chains along zigzag lines and antiferromagnetic bonds between neighboring zigzag chains, see Fig.~\ref{fig:zigzag}(a) for a case with antiferromagnetic $z$ bonds (dubbed ``$z$-zigzag''). Symmetry-equivalent zigzag states (``$x$-zigzag'' and ``$y$-zigzag'') are obtained by a $2\pi/3$ rotation about one site in real space in conjunction with a $2\pi/3$ rotation about the $[111]$ axis in spin space;\cite{footnote4} in addition there is a trivial global spin-flip (Ising) symmetry. At zero external field, any zigzag state is therefore at least six-fold degenerate.

Let us start with the case of the nearest-neighbor Heisenberg-Kitaev model with $K_1 > -J_1 > 0$, as originally proposed in Ref.~\onlinecite{chaloupka2013}. In this scenario, the zigzag chains are stabilized by a ferromagnetic Heisenberg interaction, while the antiferromagnetic ordering between neighboring zigzag chains is induced by an antiferromagnetic Kitaev interaction. In the situation with antiferromagnetically ordered $z$ bonds and ferromagnetic chains along $x$ and $y$ bonds [``$z$-zigzag state'', Fig.~\ref{fig:zigzag}(a)], the Kitaev coupling is satisfied only when the spins are aligned along the $z$ axis in spin space, $\vec S_i \parallel \pm \vec e_z$. This is illustrated in Figs.~\ref{fig:zigzag}(c) and (d). Equivalently, for the $x$- ($y$-)zigzag states, the spins point along $\pm \vec e_x$ ($\pm \vec e_y$). In Scenario 1, there are therefore six possible zigzag ground states at zero field. We denote them as ``cubic-axis zigzag.''

In general, when a collinear antiferromagnet is brought into a small external magnetic field, the field-dependent part $\mathcal H_{\vec h}$ of the energy is minimized when the staggered order is perpendicular to $\vec h$. In a Heisenberg system with $\mathrm{SU}(2)$ spin rotation invariance, such alignment is always possible already at infinitesimal field strength. For increasing field strength, all spins then cant homogeneously, i.e., with a common ``canting angle'' towards the magnetic field axis, until the system reaches saturation at some critical field strength $\hc$. In the classical limit the magnetization curve $m(h)$ is linear below $\hc$.

This situation drastically changes when the $\mathrm{SU}(2)$ spin symmetry is broken. In the present Scenario 1, with the possible zero-field spin alignment along the cubic axes only, it is impossible for the spins to align perpendicular to a small external field, unless the magnetic field is orthogonal to one of the three cubic axes $\vec e_x$, $\vec e_y$, and $\vec e_z$. At finite field strength, the spins therefore generically cant inhomogeneously towards the magnetic field with two different canting angles, and the classical magnetization curve becomes nonlinear. Furthermore, the nonuniformly canted zigzag states then compete with other states that allow for a uniform canting mechanism. This happens, for example, for a field in the out-of-plane $[111]$ direction. As we have demonstrated recently, the magnetization process for $K_1 > -2J_1 > 0$ undergoes a sequence of metamagnetic transitions, and eventually (if $K_1 > -2.5J_1$) exhibits a continuous transition from a vortex-like state with uniform canting angle towards the polarized state. On the other hand, for $-2J_1 > K_1 > -J_1 > 0$, there is a direct first-order transition from the nonuniformly canted zigzag state towards the polarized state.\cite{janssen2016}

By contrast, for an external field orthogonal to one of the cubic axes, e.g., $\vec h \parallel [\bar 110]$, the spins can align perpendicular to an infinitesimal $\vec h$. The spins therefore cant towards $\vec h$ with a uniform canting angle and a linear classical magnetization curve, until eventually a continuous transition towards the polarized phase occurs.

While the nature of the magnetization process itself is highly anisotropic in this scenario, both the \emph{magnitude} of the magnetic susceptibility at small fields and the critical field strength $\hc$ (at which the transition to the high-field state occurs) only slightly depend on the field direction. Classically, one finds, for instance,
\begin{align}
 \hc/S =
 \begin{cases}
  2K_1 - 3|J_1|, & \text{for } \vec h \parallel [111],\\
  2K_1 - 2|J_1|, & \text{for } \vec h \parallel [\bar 110],
 \end{cases}
\end{align}
as long as the high-field transition is continuous. Note that $\hc$ is even larger for the in-plane $[\bar 110]$ direction than for the out-of-plane $[111]$ direction [in contrast to the anisotropy in the field response observed in \rucl\ (Refs.~\onlinecite{johnson2015, majumder2015})].

Our numerical simulations for the prototype Parameter Set 1 are fully consistent with these expectations, see Fig.~\ref{fig:mag_set1}. For $\vec h \parallel [001]$ [Fig.~\ref{fig:mag_set1}(b)] and $\vec h \parallel [\bar 110]$ [Fig.~\ref{fig:mag_set1}(c)] the spins cant homogeneously, i.e., with a uniform canting angle, towards the magnetic field axis, resulting in a linear classical magnetization curve in the intermediate-field regime. At the critical field strength $\hc$, there is a continuous transition towards the polarized high-field phase with uniform $\langle \vec S \rangle \parallel \vec h$.
For $\vec h \parallel [111]$ [Fig.~\ref{fig:mag_set1}(a)] and $\vec h \parallel [11\bar2]$ [Fig.~\ref{fig:mag_set1}(d)], for which no spin alignment perpendicular to $\vec h$ at small field is possible, the canting becomes inhomogenous and the magnetization curve becomes nonlinear with a positive curvature in the intermediate-field regime. For this particular Parameter Set 1 with $K_1 < -2J_1$, and these field directions, there are no metamagnetic transitions, and, instead, there is a direct first-order transition towards the polarized phase. Magnetization curves for other parameter sets with $K_1 > -2J_1$, showing various novel large-unit-cell and vortex-like phases at intermediate field strengths are displayed in Ref.~\onlinecite{janssen2016}.
In the classical limit, $S \to \infty$, the high-field phase is fully polarized, $|\langle \vec S \rangle/S| = 1$, while quantum fluctuations reduce the magnetization below full saturation, $|\langle \vec S \rangle|/S < 1$, for $S=1/2$.

%%%%%%%%%%%%%%%%%%%%%%%%%%%%%%%%%%%%%%%%%%%%%%%%%%%%%%%%%%%%%%%%%%%%%%%%%%%
%%%%%%%%%%%%%%%%%%%%%%%%%%%%%%%%%%%%%%%%%%%%%%%%%%%%%%%%%%%%%%%%%%%%%%%%%%%
%%%%%%%%%%%%%%%%%%%%%%%%%%%%%%%%%%%%%%%%%%%%%%%%%%%%%%%%%%%%%%%%%%%%%%%%%%%

\begin{figure*}[t]
\includegraphics[scale=0.75]{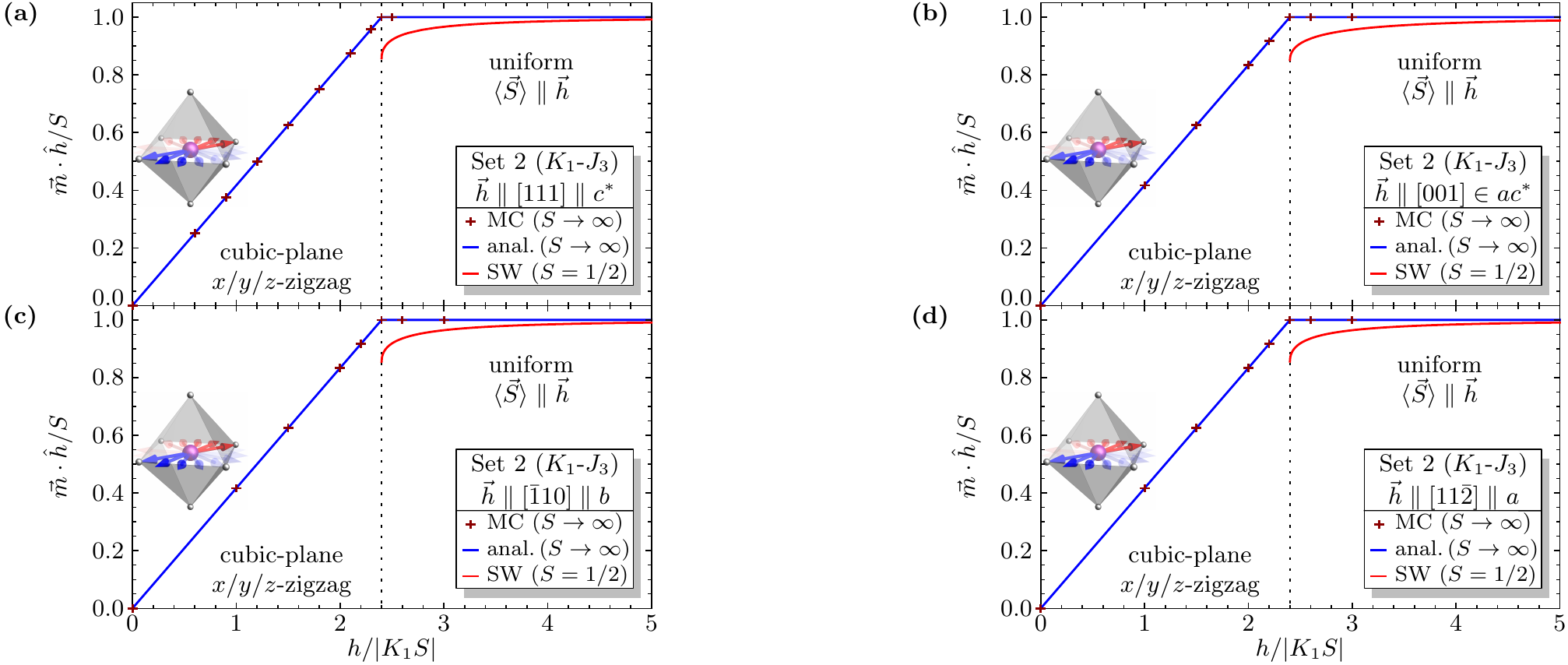}
\caption{
Magnetization in field direction for $K_1$-$J_3$ model with $K_1 < 0$ and $J_3 > 0$, using Parameter Set 2.
At zero field, the spins lie in the $xy$ ($z$-zigzag), $yz$ ($x$-zigzag), or $xz$ ($y$-zigzag) plane and can always align perpendicular to $\vec h$ upon switching on a small magnetic field in an arbitrary direction. Therefore, the classical magnetization curve is independent of the field direction in this model with $\Gamma_1 = 0$. The quantum corrections (red/light curves, high-field phases only), however, do depend on the field axis as a consequence of the order-from-disorder mechanism.
}
\label{fig:mag_set2}
\end{figure*}

\begin{figure*}[tb]
 \includegraphics[scale=0.75]{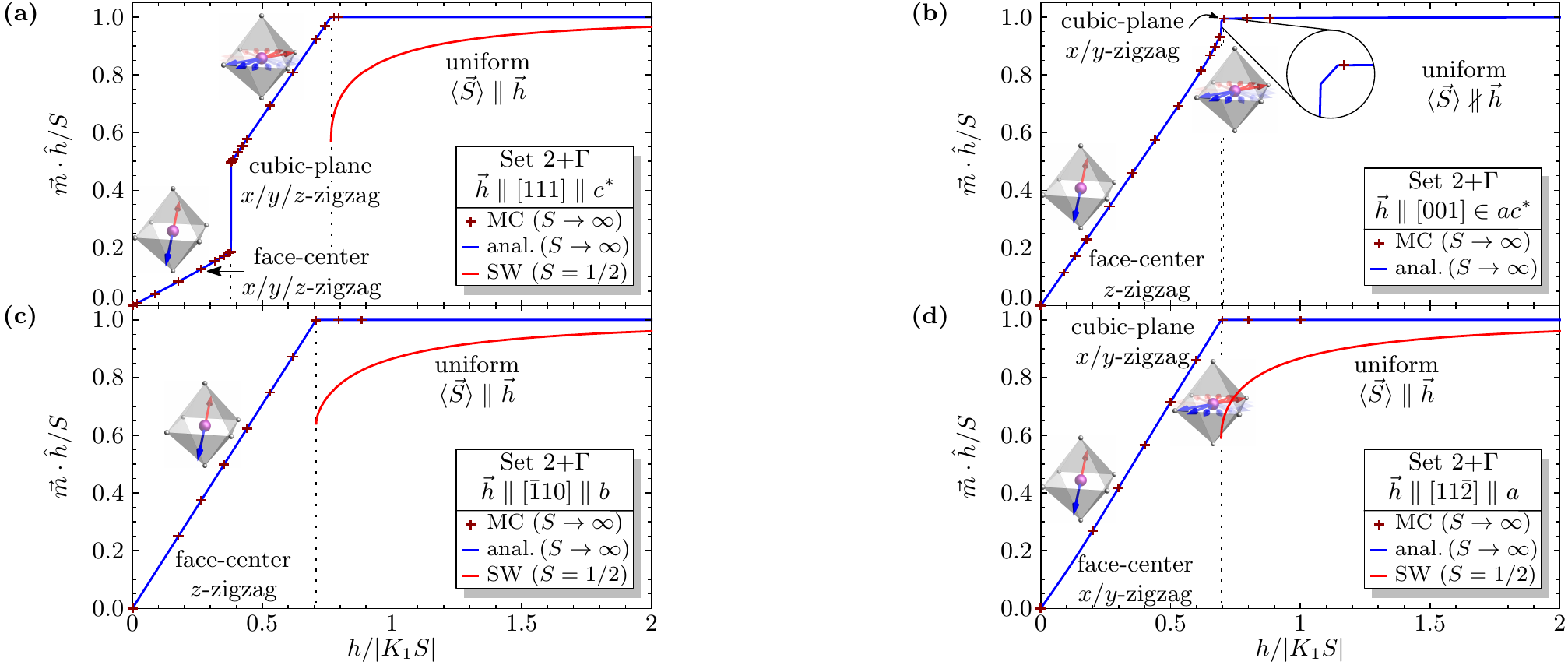}
\caption{
Magnetization in field direction for $J_1$-$K_1$-$\Gamma_1$-$J_2$-$K_2$-$J_3$ model with $K_1 < 0$, $J_3 > 0$, and small $0< \Gamma_1 \ll |K_1|$, using Parameter Set 2+$\Gamma$. At zero field, the spins align along the $[11\bar\epsilon]$, $[1\bar\epsilon1]$, or $[\bar\epsilon11]$ direction, with $0<\epsilon\ll 1$, and then cant nonuniformly towards the field for $0<h < h_{\mathrm{c}1}$ and $\vec h \parallel [111]$ (a) or $\vec h \parallel [001]$ (b). At $h_{\mathrm{c}1}$, there is a discontinuous transition towards a phase in which the spins cant uniformly, leading to a linear magnetization curve as in Fig.~\ref{fig:mag_set2}. For $\vec h \parallel [001]$ this intermediate phase is tiny, see blow-up in (b). In this case, the transition is characterized by change of the static structure factor from $z$-zigzag to $x/y$-zigzag. There are no intermediate phases for $\vec h \parallel [\bar110]$ (c). The curve for $\vec h \parallel [11\bar2]$ can be understood as a (very weak) crossover.
For all field directions, the transition transition towards the uniform high-field phase at $\hc$ is continuous.
}
\label{fig:mag_set2pg}
\end{figure*}

\section{Scenario 2: Zigzag order from ferromagnetic \texorpdfstring{$K_1$}{K1} and antiferromagnetic \texorpdfstring{$J_3$}{J3}}
\label{sec:zigzag2}

The second possible mechanism, suggested to stabilize the zigzag state in both {\rucl} and {\nio}, is characterized by a ferromagnetic Kitaev coupling $K_1$ in conjunction with a third-neighbor antiferromagnetic Heisenberg interaction $J_3$ between opposite sites of the same hexagon. In this case, the ferromagnetic zigzag chains (along, e.g., $x$ and $y$ bonds) are stabilized by the Kitaev coupling, while the antiferromagnetic ordering between neighboring zigzag chains is stabilized by $J_3$. This is illustrated in Figs.~\ref{fig:zigzag}(e) and (f). Classically, the spins in the $z$-zigzag state can therefore point anywhere in the $xy$ plane, and analogously for the $x$- and $y$-zigzag states. Consequently, the system has an accidental continuous classical ground-state degeneracy of $\mathbbm Z_3 \times \mathrm{U}(1)$. We denote these states as ``cubic-plane zigzag.''
In the real systems, the degeneracy is partially lifted by three generically competing effects:
(i) Quantum fluctuations drive an order-from-disorder mechanism which favors a spin alignment along the cubic-axes direction, i.e., $\vec S_i \parallel \pm\vec e_x$ or $\pm\vec e_y$ for the $z$-zigzag state and analogously for the $x$- and $y$-zigzag state.\cite{baskaran2008, sizyuk2016, rousochatzakis2017b}
(ii) A small off-diagonal interaction $\Gamma_1 > 0$ ($\Gamma_1 < 0$) lifts the degeneracy, e.g., in the $z$-zigzag state in favor of a state in which $\vec S_i \parallel \pm [110]$ ($\vec S_i \parallel \pm [\bar 110]$).\cite{chaloupka2015, sizyuk2016}
(iii) A small external magnetic field $\vec h$ favors a state in which the spins are aligned perpendicular to $\vec h$.

\begin{figure*}[tb]
 \includegraphics[scale=0.75]{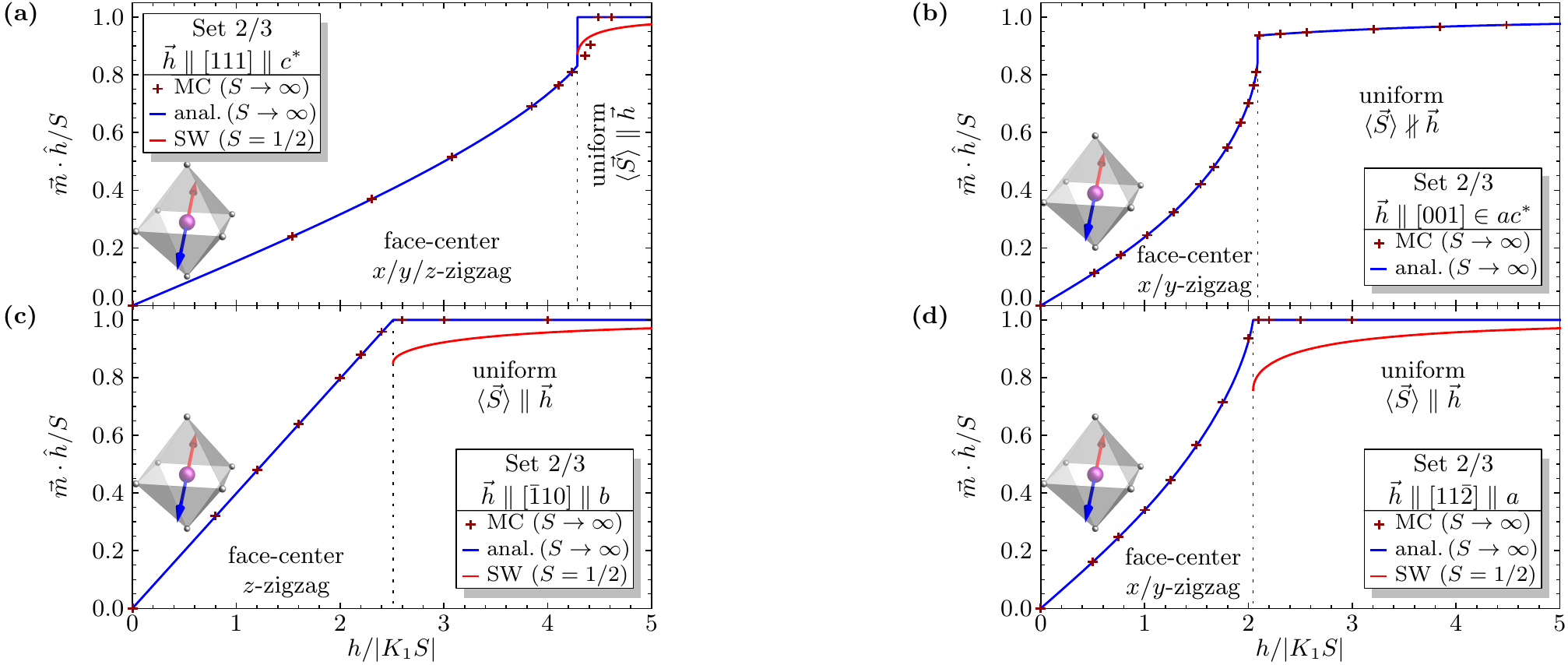}
 \caption{Magnetization in field direction for $J_1$-$K_1$-$\Gamma_1$-$J_3$ model with $K_1 < 0$, $J_3 > 0$, and large $\Gamma_1 > 0$, using Parameter Set 2/3. In contrast to Fig.~\ref{fig:mag_set2pg}, there are no intermediate phases as the ``would-be'' transition at $h_{\mathrm{c}1}$ is preempted by a first-order transition at $\hc$ towards the uniform high-field phase. For $\vec h \parallel [001]$, the high-field phase exhibits a finite transversal magnetization, leading to a nonsaturated classical magnetization $\vec m \cdot \hat h/S < 1$ for any finite $h$, see Sec.~\ref{sec:gamma}.}
\label{fig:mag_set23}
\end{figure*}

For conceptual clarity, let us first discuss the magnetization process within an idealized classical $K_1$-$J_3$ model (using, e.g., Parameter Set 2). For any field direction, the plane perpendicular to $\vec h$ intersects all three cubic planes $xy$, $yz$, and $zx$ in (at least) one axis. The zigzag spins can therefore always align perpendicular to a small external field. They cant uniformly towards the magnetic field and exhibit in the classical limit a direct continuous transition towards the polarized state in the minimal $K_1$-$J_3$ model at
\begin{equation} \label{eq:h0-set2}
 \hc/S = 6 J_3, \quad \text{for all } \vec h/h,
\end{equation}
independent of $K_1<0$. Furthermore, there is a residual $\mathbbm Z_3$ ground-state degeneracy at finite field for all field directions. This is in perfect agreement with our numerical results for Parameter Set 2, Fig.~\ref{fig:mag_set2}. Note that the presence of further Heisenberg or Kitaev interactions, such as $J_1$, $J_2$, and $K_2$ modify Eq.~\eqref{eq:h0-set2}, but do not spoil the isotropy of the magnetization process in the classical limit and when $\Gamma_1 = 0$.

Let us now include the effects of perturbations away from this idealized limit. Both quantum fluctuations and off-diagonal $\Gamma$ interactions lift the continuous ground-state degeneracy in favor of one or two preferred axes, which, as discussed above, differ among the two cases. At zero field, the spins align along the axis that corresponds to the stronger one of the two perturbations.\cite{sizyuk2016}
For a generic field direction that is not perpendicular to this axis, the zigzag spins will then cant nonuniformly towards the field axis for small fields $h \ll \max(\hbar \omega, |\Gamma_1|)$, where $\hbar \omega$ is the characteristic energy scale of the order-from-disorder effects. If both order-from-disorder effects and off-diagonal interactions are weak (e.g., for large $J_3/|K_1|$ or large $S$, and small $\Gamma_1$), there will be a transition (or crossover) at some finite field strength towards the canted version of the cubic-plane zigzag state with uniform canting angle. We confirm this expectation numerically for the Parameter Set 2+$\Gamma$ in Fig.~\ref{fig:mag_set2pg}. At zero field, the spins align along the $[11\bar\epsilon]$, $[1\bar\epsilon1]$, or $[\bar\epsilon11]$ direction, with $0<\epsilon\ll 1$. Upon switching on a small magnetic field $\vec h$ in, e.g., the $[111]$ or $[001]$ direction, they cannot align perpendicular to $\vec h$, and therefore cant nonuniformly towards it, leading to a nonlinear magnetization curve. At a critical field strength $h_{\mathrm{c}1}$, however, Figs.~\ref{fig:mag_set2pg}(a) and (b) show a transition towards an intermediate phase with a uniform canting and a linear magnetization curve. For this parameter set, there is no true transition for the two in-plane directions displayed in Figs.~\ref{fig:mag_set2pg}(c) and (d), although the latter case can be understood as a (very weak) crossover.

Hence, the zigzag state that is stabilized by a ferromagnetic $K_1$ and antiferromagnetic $J_3$, in the presence of only weak $\Gamma_1$, shows a metamagnetic transition at intermediate field strength for generic (but not all) field directions. The transition towards the polarized state, by contrast, occurs at a field strength that is, up to small corrections due to quantum fluctuations and perturbations from the off-diagonal interaction, independent of the field axis. For stronger perturbations as, for instance, in the Parameter Set 2/3 with sizable $\Gamma_1 \sim |K_1| \sim 2J_3$, the ``would-be'' transition towards the canted cubic-plane zigzag state may be preempted by a first-order transition towards the polarized state, see Fig.~\ref{fig:mag_set23}.
Consequently, the transition towards the high-field phase at $\hc$ is discontinuous for $\vec h \parallel [111]$ [Fig.~\ref{fig:mag_set23}(a)] and $\vec h \parallel [001]$ (b), but remains continuous for $\vec h \parallel [\bar110]$ (c) and $\vec h \parallel [11\bar2]$ (d).

%%%%%%%%%%%%%%%%%%%%%%%%%%%%%%%%%%%%%%%%%%%%%%%%%%%%%%%%%%%%%%%%%%%%%%%%%%%
%%%%%%%%%%%%%%%%%%%%%%%%%%%%%%%%%%%%%%%%%%%%%%%%%%%%%%%%%%%%%%%%%%%%%%%%%%%
%%%%%%%%%%%%%%%%%%%%%%%%%%%%%%%%%%%%%%%%%%%%%%%%%%%%%%%%%%%%%%%%%%%%%%%%%%%

\section{Scenario 3: Zigzag order from positive \texorpdfstring{$\Gamma_1$}{Gamma1} and ferromagnetic \texorpdfstring{$J_1$}{J1}, \texorpdfstring{$K_1$}{K1}}
\label{sec:zigzag3}

\begin{figure*}[tb]
 \includegraphics[scale=0.75]{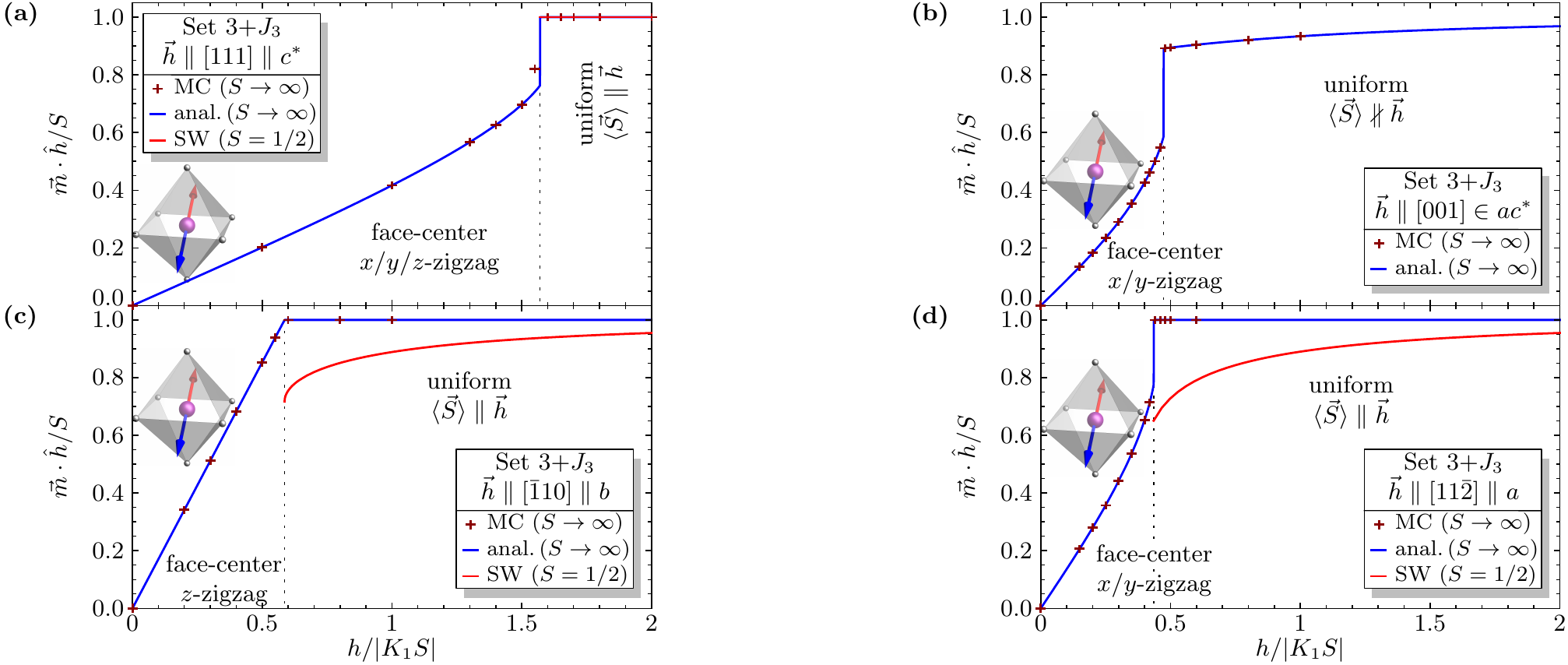}
 \caption{Magnetization in field direction for $J_1$-$K_1$-$\Gamma_1$-$J_3$ model with $J_1 < 0$, $K_1 < 0$, $J_3 > 0$, and large $\Gamma_1 > 0$ using Parameter Set 3+$J_3$. The critical field $\hc$ is strongly anisotropic, with the largest $\hc$ among the four shown here occurring for the out-of-plane direction $\vec h \parallel [111]$ (a). For this particular field direction, the leading-order quantum correction to the classical magnetization happens to vanish due to an accidental cancellation in linear spin-wave theory (red/light curves, high-field phases only). For the intermediate field direction $\vec h \parallel [001]$, the uniform high-field phase is characterized by a finite transversal magnetization, $\langle \vec S \rangle \nparallel \vec h$, leading to a nonsaturated magnetization in field direction already on the classical level (b). The transition at $\hc$ is typically discontinuous, except for $\vec h \parallel [\bar 110]$, which corresponds to a field direction that is perpendicular to the zero-field moments (c).}
\label{fig:mag_set3pj3}
\end{figure*}

Finally, we consider the zigzag state that is stabilized within a nearest-neighbor model with strong positive off-diagonal interaction $\Gamma_1$.
We emphasize that the classical ground state of the plain $K_1$-$\Gamma_1$ model with $K_1 < 0$ and $\Gamma_1> 0$ is a multi-$\vec Q$ state with six major and three minor Bragg peaks in the first Brillouin zone. The zigzag ground state can, however, be recovered for nearest-neighbor interactions only, when the model is supplemented with a ferromagnetic Heisenberg interaction $J_1$, provided that the ratio $\Gamma_1 / |K_1|$ is large enough. This is demonstrated in detail in the appendix. The minimal description of Scenario 3 is therefore given by a $J_1$-$K_1$-$\Gamma_1$ model.
In this model, the spins in, e.g., the $z$-zigzag state are aligned for zero external field along an $[xxz]$ direction determined by $z/x = f(\Gamma_1/|K_1|)$, with
\begin{align} \label{eq:ordered-moments}
 f(\zeta) = \frac{2 + \zeta - \sqrt{4 + 4\zeta + 9\zeta^2}}{2\zeta}, \qquad \zeta = \Gamma_1 / |K_1|,
\end{align}
and thus $-1\leq z/x \leq 0$. Note that the value of the Heisenberg interaction $J_1$ does not affect the direction of the ordered moments at zero field. The same applies to further isotropic interactions ($J_2$, $J_3$), such that Eq.~\eqref{eq:ordered-moments} is valid for all parameter sets in Table~\ref{tab:par} with $K_1 < 0$, $\Gamma_1 >0$, and $K_2 = 0$.
For small $\Gamma_1 \ll |K_1|$, we have $z/x \to 0$, such that $\vec S_i \parallel \pm [110]$. For large $\Gamma_1 \gg |K_1|$, on the other hand, one finds $z/x \to -1$, and hence $\vec S_i \parallel \pm [11\bar 1]$. In this latter case, the effective moments in \rucl\ and \nio\ would point towards the centers of opposite faces of the surrounding chlorine or oxygen octahedra. This is illustrated in Figs.~\ref{fig:zigzag}(g) and (h).
To stay within a sufficiently simple naming scheme, we therefore denote these states, in a slight abuse of notation for finite $\Gamma_1/|K_1|$, as ``face-center zigzag.''
In the $x$- and $y$-zigzag states, the spins are analogously aligned along an $[xyy]$ and $[xyx]$ direction. The zero-field ground state in this scenario is therefore six-fold degenerate.

Due to the alignment of the zigzag spins at zero field, the magnetization process crucially depends on the direction of the external field. If the magnetic field is perpendicular to the zero-field ordered-moment direction, a simple canting mechanism with uniform canting angle and, in the classical limit, a linear magnetization curve can occur. This happens, for example, for the in-plane direction $\vec h \parallel [\bar 110]$. At a critical field strength $\hc$ with (in the simplest $J_1$-$K_1$-$\Gamma_1$ model)
\begin{align}
 \hc/S & = 2 J_1 + K_1 - \tfrac{1}{2}\Gamma_1
 \nonumber \\ & \quad + \sqrt{K_1^2 - K_1 \Gamma_1 + \tfrac{9}{4} \Gamma_1^2},
 \qquad \text{for } \vec h \parallel [\bar 110],
\end{align}
there is a direct and continuous transition towards the polarized state.
There is another linearly independent direction of the form $[xxz']$, perpendicular to the zero-field $[xxz]$ moments in the $z$-zigzag state. A uniform canting and a continuous high-field transition are expected in this field direction as well. Analogous perpendicular directions $[x'yy]$ and $[xy'x]$ exist for the $x$- and $y$-zigzag states. All these directions, however, depend on the ratio $\Gamma/K_1$ via the function $f$.

For other field directions, however, the zigzag spins cannot align perpendicular to the field axis for small $h$. Consequently, they cant with two different angles towards the magnetic field. We demonstrate numerically in Figs.~\ref{fig:mag_set3pj3}(a), (b), and (d) that the magnetization curve is nonlinear and the transition to the high-field state for these field directions is typically first order. By contrast, the magnetization curve is linear and the high-field transition is continuous for $\vec h \parallel [\bar 110]$ [Figs.~\ref{fig:mag_set3pj3}(c)]. Fig.~\ref{fig:mag_set3pj3} also shows that the critical field strength $\hc$, at which the transition to the high-field phase occurs, now critically depends on the field axis, with the largest $\hc$ among the four shown here for the out-of-plane direction $\vec h \parallel [111]$ [Fig.~\ref{fig:mag_set3pj3}(a)]. This is in sharp contrast to the situation with small $|\Gamma_1| \ll |K_1|$ shown in Figs.~\ref{fig:mag_set1}--\ref{fig:mag_set2pg}. Furthermore, the high-field phase for $h>\hc$ shown in Fig.~\ref{fig:mag_set3pj3}(b) for $\vec h \parallel [001]$ is characterized by a nonsaturated classical magnetization in field direction, corresponding to a finite transversal magnetization perpendicular to $\vec h$. We elaborate on these key aspects of the models with strong $\Gamma_1$ in the following section.

%%%%%%%%%%%%%%%%%%%%%%%%%%%%%%%%%%%%%%%%%%%%%%%%%%%%%%%%%%%%%%%%%%%%%%%%%%%
%%%%%%%%%%%%%%%%%%%%%%%%%%%%%%%%%%%%%%%%%%%%%%%%%%%%%%%%%%%%%%%%%%%%%%%%%%%
%%%%%%%%%%%%%%%%%%%%%%%%%%%%%%%%%%%%%%%%%%%%%%%%%%%%%%%%%%%%%%%%%%%%%%%%%%%

\section{Consequences of strong \texorpdfstring{$\Gamma_1$}{Gamma1}}
\label{sec:gamma}

The presence of a sizable off-diagonal interaction $\Gamma_1$ as suggested within Scenario~3 has two characteristic consequences, which are in sharp contrast to the effects of the Heisenberg and Kitaev interactions and which allow to test this scenario in future experiments.

\subsection{Anisotropic magnetic response}

First, we note that identifying the sign of $\Gamma_1$ as ``ferromagnetic'' or ``antiferromagnetic'' is inappropriate. The nature of the off-diagonal interaction for fixed sign of $\Gamma_1$ in fact \emph{depends} on the direction of the ordered moments.
Let us assume for simplicity a uniform spin state $\vec S_i \equiv \vec S$ in an arbitrary direction, as it would occur for a very strong magnetic field. The argument, however, can be made in a similar way also for components of the spins in the direction of a small external field. We find it convenient to choose the new orthonormal basis $\{\vec e_1, \vec e_2, \vec e_3\}$ in which the coordinate axes are aligned along the crystallographic $a$, $b$, and $c^*$ axes in Fig.~\ref{fig:zigzag}(b), i.e., $\vec e_1 \propto [11\bar2]$, $\vec e_2 \propto [\bar110]$, and $\vec e_3 \propto [111]$.
The uniform state can then be written as
\begin{equation}
 \vec S/S = \sin \vartheta \cos \varphi \,\vec e_1 + \sin \vartheta \sin \varphi \,\vec e_2 + \cos \vartheta \,\vec e_3,
\end{equation}
where $\vartheta$ and $\varphi$ are the spherical angles in the $\{\vec e_1, \vec e_2, \vec e_3\}$ basis, such that $\vartheta = 0\text\textdegree$ corresponds to the $c^*$ axes and $\vartheta = 90\text\textdegree$ corresponds to the $ab$ plane.
In this basis, the field-independent part of the total classical energy becomes particularly simple,
\begin{align} \label{eq:ergall}
  \langle E_0 / (NS^2) \rangle_{\vec S_i \equiv \vec S} & = \frac{3J_1 + K_1 + 6J_2 +2 K_2 + 3J_3}{2}
  \nonumber \\ & \quad
  + \frac{\Gamma_1}{4}(1+3\cos 2\vartheta).
\end{align}
Notably, the energy becomes independent of the azimuthal angle $\varphi$, and only the off-diagonal $\Gamma_1$ interaction term depends on the polar angle $\vartheta$.
For any in-plane direction, this anisotropic part becomes $\langle E_0 / (NS^2) \rangle_{\Gamma_1} = -\Gamma_1/2 < 0$, while $\langle E_0 / (NS^2) \rangle_{\Gamma_1} = +\Gamma_1 > 0$ for the out-of-plane direction $c^*$. In general, the $\Gamma_1$-dependent contribution is negative for $55\text{\textdegree}  < \vartheta < 125\text{\textdegree}$ and positive otherwise (assuming that $\Gamma_1 > 0$).
Within the honeycomb plane, a positive $\Gamma_1$ therefore favors a \emph{ferromagnetic} spin alignment over an antiferromagnetic one. If the spins are forced to lie perpendicular to the honeycomb plane, on the other hand, a positive $\Gamma_1$ favors an \emph{antiferromagnetic} spin pattern over a ferromagnetic alignment.

\begin{figure*}[tb]
 \includegraphics[width=0.47\textwidth]{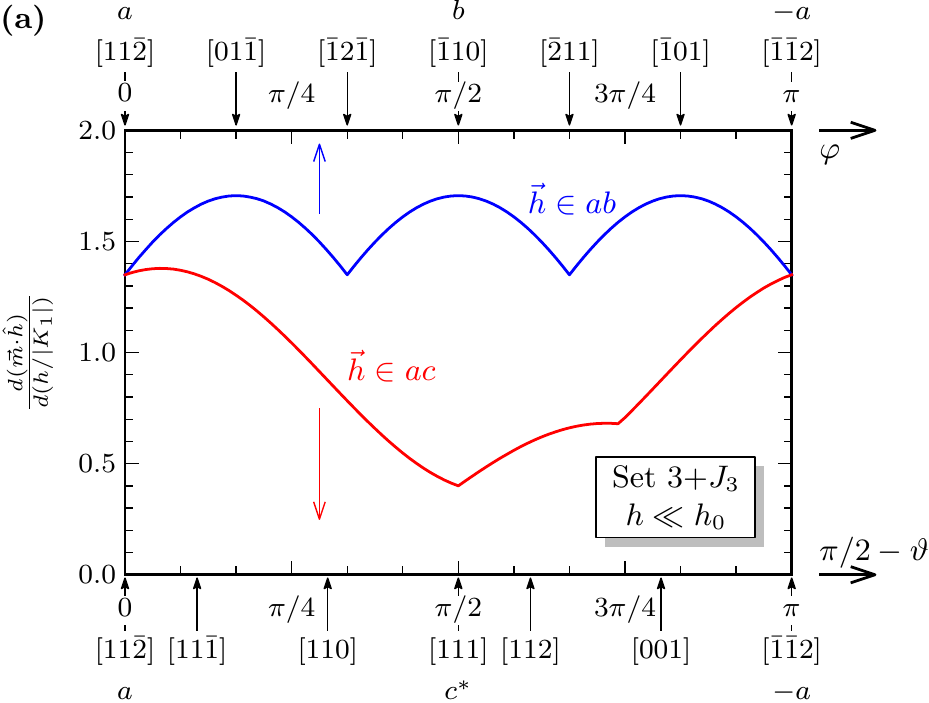}\hfill
 \includegraphics[width=0.47\textwidth]{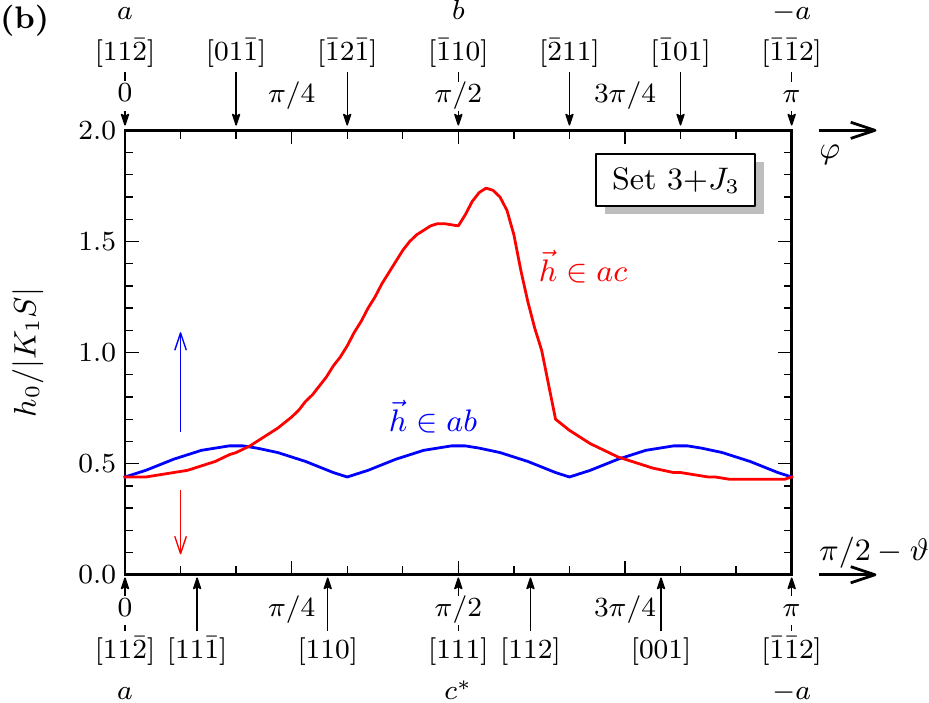}
 \caption{{(a)} Susceptibility $\chi(\vartheta,\varphi) = d (\vec m \cdot \hat h) / d(h/|K_1|)$ at zero temperature and in the classical limit as function of magnetic field axis $\vec h/h$ for Parameter Set 3+$J_3$ with sizable $\Gamma_1 > 0$. $\vartheta$ and $\varphi$ are the spherical angles in the $a,b,c^*$ basis, such that $\vartheta = 0$ corresponds to the crystallographic $c^*$ axis perpendicular to the honeycomb layer, denoted by $[111]$. $\vartheta = \pi/2$ with $\varphi=0$ ($\varphi = \pi/2$) corresponds to the in-plane crystallographic $a$ ($b$) axis, denoted by $[11\bar2]$ ($[\bar110]$). Red (lower) curve: Out-of-plane susceptibility $\chi(\vartheta,0)$ from $a$ to $c^*$ to $-a$. Blue (upper) curve: In-plane susceptibility $\chi(0, \varphi)$ from $a$ to $b$ to $-a$. {(b)} Same as (a), but now showing the critical field strength $\hc$ of the transition towards the high-field phase.
} \label{fig:susceptibility-h0}
\end{figure*}

Consequently, the response of a system with a large off-diagonal interaction $\Gamma_1 > 0$ to an external magnetic field crucially depends on the field axis: For an out-of-plane direction, the off-diagonal interaction has an antiferromagnetic character. Increasing $\Gamma_1 > 0$ lifts the energy of the polarized state, leads to an increase of the critical field strength $\hc$ at which the transition to the high-field state occurs, and makes the system less susceptible to the external field.
For an in-plane direction, by contrast, a positive $\Gamma_1 > 0$ behaves as if the interaction were ferromagnetic. Increasing $\Gamma_1$ reduces the critical field strength, and makes the system \emph{more} susceptible to the external field.
We illustrate this behavior for the low-temperature susceptibility $\chi = d(\vec m \cdot \hat h)/d(h/|K_1|)$ at small in-plane and out-of-plane fields $h \ll \hc$ as well as the critical field strength $\hc$ in Fig.~\ref{fig:susceptibility-h0}. Here, we have used Parameter Set 3+$J_3$ as an example, and have solved the finite-field problem in the classical limit by employing the methods described in Sec.~\ref{sec:methods}. Without any $g$-factor anisotropy, the ratio between in-plane and out-of-plane susceptibility for this parameter set would be $\chi_{a}/\chi_{c^*} \simeq 3.3$, and the ratio of critical field strengths would be $h_{0,a}/h_{0,c^*} \simeq 0.28$. Between different in-plane directions, on the other hand, the anisotropy is much smaller, e.g., $h_{0,a}/h_{0,b} \simeq 0.76$ for the ratio of the critical field strengths.
Note that a negative $\Gamma_1 < 0$ has precisely the opposite effect: It leads to a ferromagnetic character of the off-diagonal interaction for an out-of-plane magnetic field $\vec h \parallel c^*$ and an antiferromagnet character for an in-plane field $\vec h \in ab$.
Hence, a positive off-diagonal interaction $\Gamma_1 > 0$ naturally explains the strong magnetic anisotropy observed experimentally in \rucl.\cite{footnote6} This is further discussed in Sec.~\ref{sec:discussion}.

\subsection{High-field transversal magnetization}

Second, we investigate the nature of the high-field phase in the presence of a finite $\Gamma_1$. To this end, it is instructive to recall the symmetries of our (idealized cubic) system.\cite{jiang2011}
Due to the bond-dependent interactions, reflecting strong spin-orbit coupling, the continuous $\mathrm{SU}(2)$ spin rotation symmetry known from isotropic Heisenberg models is replaced by a residual discrete $C_3^*$ symmetry of $2\pi/3$ rotation around the $[111]$ axis. This corresponds to the cyclic permutation $(S_x,S_y,S_z) \mapsto (S_y,S_z,S_x)$ in spin space together with a $2\pi/3$ rotation about one site in real space.
There is also a discrete $C_2^*$ rotation symmetry about the in-plane $[\bar 110]$ axis with angle $\pi$, which corresponds to $(S_x,S_y,S_z) \mapsto (-S_y,-S_x,-S_z)$ in spin space and a reflection in real space.
Furthermore, there is a reflection symmetry $C_s^*$ through the $ac$ plane perpendicular to the $[\bar 110]$ axis, corresponding to the spin transformation $(S_x,S_y,S_z) \mapsto (S_y,S_x,S_z)$ and a reflection in real space.
The composition $C_2^* C_s^*$ maps a vector $\vec S$ in spin space onto its inverse, $\vec S \mapsto - \vec S$, which corresponds to the global spin-flip (Ising) symmetry mentioned earlier.
However, if the off-diagonal interaction vanishes, $\Gamma_1 = 0$, the model has an \emph{enhanced} symmetry which is not protected by the lattice point group: It is given by the individual inversion of components of $\vec S$, i.e., $S_x \mapsto -S_x$, $S_y \mapsto -S_y$, or $S_z \mapsto - S_z$.

An external magnetic field breaks part of the symmetry. For instance, a field in the $[001]$ direction breaks $C_3^*$ and $C_2^*$, but leaves $C_s^*$ invariant. If $\Gamma_1 = 0$, inverting the $x$ and $y$ components of $\vec S$ individually remains a symmetry, but not for $\Gamma_1 \neq 0$. At high fields $h$ beyond a certain critical field strength $\hc$, one expects a ground state that is adiabatically connected the fully polarized state at $h \to \infty$, without spontaneous symmetry breaking. Consequently, without the off-diagonal $\Gamma_1$ interaction, the spin expectation value $\langle \vec S \rangle$ in the high-field state is locked to the field axis,
\begin{align}
 \langle \vec S_i \rangle \equiv \langle \vec S \rangle \parallel \vec h, \qquad \text{for } \Gamma_1 = 0.
\end{align}
As a side note, we recall that even in this case the state at $\hc < h < \infty$ is \emph{not} fully polarized, $|\langle \vec S \rangle|/S < 1$, since quantum fluctuations reduce the magnetization as a consequence of the broken $\mathrm{SU}(2)$ spin symmetry, possibly even substantially so, see the red/light curves in Figs.~\ref{fig:mag_set1}--\ref{fig:mag_set3pj3}, as well as Ref.~\onlinecite{janssen2016}.

In the presence of a finite off-diagonal interaction $\Gamma_1 \neq 0$, however, there is no symmetry which forbids a high-field state that is \emph{not} parallel to $\vec h$, such as, e.g., $\langle \vec S_i \rangle \parallel [xxz]$ with $x/z \neq 0$ for $\vec h \parallel [001]$. If this indeed happened, it would mean that the magnetization $\vec m = \vec m_\parallel + \vec m_\perp$ would exhibit a transversal component $\vec m_\perp \neq 0$ which is perpendicular to the external field for all finite $h < \infty$, and only in the strict limit $h \to \infty$ would the total magnetization $\vec m$ become parallel to the external field $\vec h$.
In order to examine for which parameters this indeed happens in our model, and if so, for which field directions, we again consider a uniform classical high-field state $\vec S_i \equiv \vec S$ with the field-independent part of the total energy given in Eq.~\eqref{eq:ergall}.
We first note that $\langle E_0/ (NS^2) \rangle$ becomes entirely independent of $\vartheta$ and $\varphi$ when $\Gamma_1 = 0$. This is consistent with our symmetry-reasoned expectation formulated above, stating that $\langle \vec S \rangle$ is then parallel to the magnetic field $\vec h$ for all field directions.
With the off-diagonal interaction present, however, whether or not $\langle \vec S \rangle$ and $\vec h$ are collinear depends on the magnetic-field direction and the ratio $\Gamma_1/(h/S)$. In any case, since $\langle E_0/ (NS^2) \rangle$ is independent of $\varphi$, $\langle \vec S \rangle$ will always lie in the plane spanned by $\vec e_3 \parallel c^*$ and $\vec h$.
Let us denote the polar angle of $\vec h$ in the $\{\vec e_1, \vec e_2, \vec e_3\}$ basis by $\theta_h = \arccos(\vec e_3 \cdot \vec h/h)$. The angle between $\vec h$ and $\langle \vec S \rangle$ is then given by $|\theta_h - \vartheta|$. To see whether $|\theta_h - \vartheta|$ has a finite expectation value in the high-field state, we have to minimize the total energy $\langle E\rangle_{\vec S_i \equiv \vec S} = \langle E_0 + E_{\vec h} \rangle_{\vec S_i \equiv \vec S}$, where
\begin{align}
 \langle E_{\vec h}/ (NS^2) \rangle_{\vec S_i \equiv \vec S} = - \frac{h}{S} \cos(\theta_h - \vartheta).
\end{align}
This yields a conditional equation for the energy-minimizing $\vartheta = \vartheta_\text{min}$ as function of  $\theta_h \in [0,\pi]$, reading
\begin{align} \label{eq:vartheta0}
 \frac{3}{2} \Gamma_1 \sin 2 \vartheta_\text{min} + \frac{h}{S} \sin(\theta_h - \vartheta_\text{min}) = 0,
\end{align}
and which can be solved exactly for any given $\theta_h$. From this we find $\vartheta_\text{min} = 0$ for $\theta_h = \pi/2$ or $\theta_h = 0$ and large enough $(h/S)/\Gamma_1$. For $0 < \theta < \pi/2$, by contrast, $\vartheta_\text{min}$ is finite \emph{for all} $h < \infty$. Consequently, we find a finite transversal magnetization $\vec m_\perp \perp \vec h$ in the high-field state for all field directions $\vec h$ which are neither parallel nor perpendicular to the $[111]$ axis. In the classical limit, the relative size of the transversal magnetization is given by
\begin{align} \label{eq:mtrans}
 \frac{|\vec m_\perp|}{|\vec m_\parallel|} =
 \begin{cases}
  0, 		& \text{for } \vec h \perp c^*, \\
  \tan|\theta_h - \vartheta_\text{min}|, 	
		& \text{for } 0 < \theta_h \equiv \measuredangle(\vec h, c^*) < \frac{\pi}{2}, \\
		%\text{ between $ab$ and $c^*$}, \\
% 		\nparallel c^* \land \vec h \not\perp c^*, \\
  0,		& \text{for } \vec h \parallel c^*,
 \end{cases}
\end{align}
with $\vartheta_\text{min}$ determined through Eq.~\eqref{eq:vartheta0}. These results are consistent with the numerical data presented in Figs.~\ref{fig:mag_set1}--\ref{fig:mag_set3pj3}: The uniform high-field phase is characterized by a finite transversal magnetization (indicated by a nonsaturated classical magnetization in field direction) if and only if $\Gamma_1 \neq 0$ and for an external field $\vec h$ that is neither parallel nor perpendicular to the honeycomb plane. We have explicitly checked that our numerical results agree with Eq.~\eqref{eq:mtrans} also quantitatively. For instance, at $h=\hc^+$ in the $[001]$ direction (i.e., $\theta_h = 55\text\textdegree$), we find ${|\vec m_\perp|}/{|\vec m_\parallel|} \simeq 0.38$ and ${|\vec m_\perp|}/{|\vec m_\parallel|} \simeq 0.51$ for the Parameter Sets 2/3 and 3+$J_3$, respectively.

%%%%%%%%%%%%%%%%%%%%%%%%%%%%%%%%%%%%%%%%%%%%%%%%%%%%%%%%%%%%%%%%%%%%
%%%%%%%%%%%%%%%%%%%%%%%%%%%%%%%%%%%%%%%%%%%%%%%%%%%%%%%%%%%%%%%%%%%%
%%%%%%%%%%%%%%%%%%%%%%%%%%%%%%%%%%%%%%%%%%%%%%%%%%%%%%%%%%%%%%%%%%%%

\section{Discussion}
\label{sec:discussion}

\subsection{Susceptibility anisotropy}

\rucl\ (Ref.~\onlinecite{kubota2015}) and, to a somewhat lesser extent, \nio\ (Ref.~\onlinecite{singh2010}) show a distinguished anisotropy in the magnetic response. The in-plane magnetic susceptibility $\chi_{ab}$ in \rucl\ at low fields and temperatures is about four times larger than its out-of-plane counterpart $\chi_{{c}^*}$.\cite{kubota2015}
Similarly, at a field strength of, e.g., 15\,T, the magnetization is about 5--6 times larger when the external field is applied within the $ab$ plane as compared to when $\vec h \perp ab$.\cite{johnson2015}
Given that the trigonal distortion is expected to be relatively small,\cite{cao2016, park2016, tjeng2017} it appears quite unnatural to attribute such large anisotropy exclusively to the effects of an anisotropic $g$ tensor. In fact, \emph{ab-initio} calculations rather point to a fairly isotropic $g$ tensor,\cite{majumder2015, tjeng2017} suggesting an \emph{intrinsic} mechanism originating from bond-dependent spin-spin exchange interactions.

In this work, we have demonstrated that a model with a strong off-diagonal interaction $\Gamma_1 > 0$ naturally leads to a strongly anisotropic field response, without the need to assume a largely anisotropic $g$ tensor. For the particular Parameter Set 3+$J_3$, the ratio of susceptibilities between an in-plane field direction $\vec h \in ab$ and the out-of-plane direction $\vec h \parallel c^*$ becomes of the order of $\chi_{ab}/\chi_{c^*} \sim 3\text{--}4$ (depending on the particular in-plane direction chosen), which agrees with the experimental value for \rucl. We therefore suggest that the off-diagonal interaction $\Gamma_1$ in \rucl\ is positive and large, possibly of the order of $\Gamma_1 / |K_1| \sim 0.5$, as given by Parameter Set 3+$J_3$. As an aside, we note that the direction of the ordered moments at zero field as given by Eq.~\eqref{eq:ordered-moments} for $\Gamma_1 / |K_1| \sim 0.5$ follows as $\vec S/S \sim \pm[0.68,0.68,-0.26]$, which corresponds to an axis in the $ac$ plane that is tilted $\sim 40\text{\textdegree}$ away from the $a$ axis towards the $c^*$ axis. Remarkably, this appears to be in fair agreement with one of the two possible options expected for \rucl.\cite{cao2016}

The anisotropy \emph{within} the honeycomb plane is much smaller, but finite, with a characteristic $\pi/3$ periodicity resulting from the discrete $C_3^*$ lattice rotation symmetry. We predict that the in-plane susceptibility in both \rucl\ and \nio\ is maximal for a field direction along Ru-Ru bonds and Ir-Ir bonds, respectively (i.e., for $\vec h$ being parallel to the crystallographic $b$ axis in our conventions), and minimal for the in-plane direction that is perpendicular to this axis ($a$ axis in our conventions).

\subsection{Field-induced transitions}

Recent experiments report a field-induced transition in \rucl\ from a canted ordered phase into a putative quantum disordered phase when a field of around 7\,T (10\,T) is applied along an in-plane\cite{leahy2017, sears2017, zheng2017, hentrich2017, banerjee2017} (intermediate\cite{baek2017}) direction. While at least one, \cite{wolter2017} or more, \cite{banerjee2017} field-induced phase transitions appear to be observable independent of the specific sample, the nature of the adjacent phases has so far not been established beyond doubt.

We have argued that the magnetization process differentiates between the three scenarios for stabilizing a zigzag state. While the transition towards the uniform high-field phase is generically continuous when the field is directed along a Ru-Ru bond (e.g., along the $b$ axis), this is different
when the field is applied along an in-plane direction orthogonal to a Ru-Ru bond (e.g., $a$ axis): Here, the high-field transition is continuous if and only if $K_1$ is ferromagnetic and $\Gamma_1/|J_1|$ is not too large, with the Parameter Set 2/3 appearing to describe a rough upper bound for the value of the off-diagonal interaction. For larger values of $\Gamma_1/|J_1|$, such as in the Parameter Set 3+$J_3$, the transition becomes discontinuous. When the actual nature of the Kitaev interaction in \rucl\ is antiferromagnetic, as suggested within Scenario 1, the transition from the canted zigzag state into the high-field state becomes continuous only if the magnetic field lies within one of the three cubic planes, which in \rucl\ are the planes perpendicular to the Ru-Cl bonds. Detailed magnetization measurements, using high (pulsed) magnetic fields, for different field directions are clearly called for.

Our results feature phase transitions at intermediate fields only for Parameter Set 2+$\Gamma$, there between different canted states, and possibly in Scenario 1, there towards more complicated vortex and multi-$Q$ states.\cite{janssen2016} This situation might change upon accounting for (i) quantum effects beyond the semiclassical limit (possibly leading to field-induced spin-liquid behavior) and (ii) deviations from the idealized cubic structure. Investigations in these directions are left for future work.

Recent neutron diffraction measurements in \rucl\ in small magnetic field $\vec h$ applied along an in-plane direction that is orthogonal to a Ru-Ru bond (e.g., along the $a$ axis in our notation), find a depopulation of zigzag domains with ordering wavevectors $\vec Q \perp \vec h$.\cite{sears2017, banerjee2017} This observation is consistent with our calculations in the models with $\Gamma_1 > 0$ (Scenario 3) or $K_1 > 0$ (Scenario 1): A magnetic field along the in-plane $[11\bar2]$ direction, which is perpendicular to a $z$ bond in our conventions, favors the $x$- and $y$-zigzag states to the detriment of the $z$-zigzag state.

\subsection{Transversal magnetization at high fields}

We have furthermore shown that the off-diagonal $\Gamma_1$ interaction leads to a finite transversal magnetization in the uniform high-field phase if the magnetic field axis is neither parallel nor perpendicular to the honeycomb plane. As the off-diagonal interaction already breaks part of the symmetry of the original Heisenberg-Kitaev model explicitly, this is possible without spontaneous symmetry breaking. For a magnetic field along Ru-Cl bonds, for instance, the transversal magnetization becomes comparatively large and should in principle be observable in \rucl. Its magnitude is a direct measure of the size of $\Gamma_1$, see Eqs.~\eqref{eq:vartheta0} and \eqref{eq:mtrans}.

\section{Conclusions and outlook}

In this paper, we have studied the magnetic-field behavior of various extended Heisenberg-Kitaev models on the honeycomb lattice. The magnetization processes sensitively depend on the signs and relative sizes of the model parameters, and we have assessed the validity of the models in light of the experimental observations in \rucl\ and \nio. We have also predicted a number of further measurable consequences.

With these theoretical predictions, we believe that a detailed experimental characterization of the magnetization processes of the zigzag states should allow to fix the sign of the Kitaev interaction $K_1$ and determine the magnitude of the off-diagonal interaction $\Gamma_1$ in \rucl\ and \nio.
Once an appropriate effective spin Hamiltonian has been established, a full quantum computation that goes beyond the present semiclassical analysis is indispensable. This should also allow to understand and characterize the field-induced phases found experimentally in \rucl.\cite{leahy2017,baek2017,sears2017,zheng2017,hentrich2017,banerjee2017}

%%%%%%%%%%%%%%%%%%%%%%%%%%%%%%%%%%%%%%%%%%%%%%%%%%%%%%%%%%%%%%%%%%%%

\acknowledgments

We thank B. B\"uchner, P. Lampen-Kelley, R. Moessner, S. Nagler, N. Perkins, S. Rachel, R. Valent\'i, J. van den Brink, A. Vishwanath, Z. Wang, S. Winter, and A. Wolter
for illuminating discussions and collaboration on related work.
This research was supported by the DFG through SFB 1143 and GRK 1621. ECA was supported by FAPESP (Brazil) Grant No.\ 2013/00681-8 and CNPq (Brazil) Grant No.\ 302065/2016-4.

%%%%%%%%%%%%%%%%%%%%%%%%%%%%%%%%%%%%%%%%%%%%%%%%%%%%%%%%%%%%%%%%%%%%
%%%%%%%%%%%%%%%%%%%%%%%%%%%%%%%%%%%%%%%%%%%%%%%%%%%%%%%%%%%%%%%%%%%%
%%%%%%%%%%%%%%%%%%%%%%%%%%%%%%%%%%%%%%%%%%%%%%%%%%%%%%%%%%%%%%%%%%%%

\appendix

\section{Zigzag state in nearest-neigbor \texorpdfstring{$J_1$-$K_1$-$\Gamma_1$}{J1-K1-Gamma1} model}
\label{app:zigzag3}

Recently, a minimal nearest-neighbor $K_1$-$\Gamma_1$ model with $K_1 < 0$ and $\Gamma_1 > 0$ was suggested to describe the physics of \rucl.\cite{ran2017, wang2016} In this appendix, we demonstrate that such an oversimplified model---without any Heisenberg-type interactions---is (at least in the classical limit) \emph{not} sufficient to stabilize a zigzag antiferromagnet when $K_1 < 0$ and $\Gamma_1 > 0$. A finite ferromagnetic Heisenberg interaction $J_1 < 0$ and a large enough ratio $\Gamma_1/|K_1|$ are needed to recover the zigzag ground state. This result is in contrast to the exploratory study of Ref.~\onlinecite{rau2014}, which used a combined Luttinger-Tisza and single-$\vec Q$ analysis for the classical energy minimization. Here, we employ classical Monte-Carlo simulations which are capable to also detect multi-$\vec Q$ and incommensurate phases in a sufficiently unbiased way.

\begin{figure}[b]
\includegraphics[width=\linewidth]{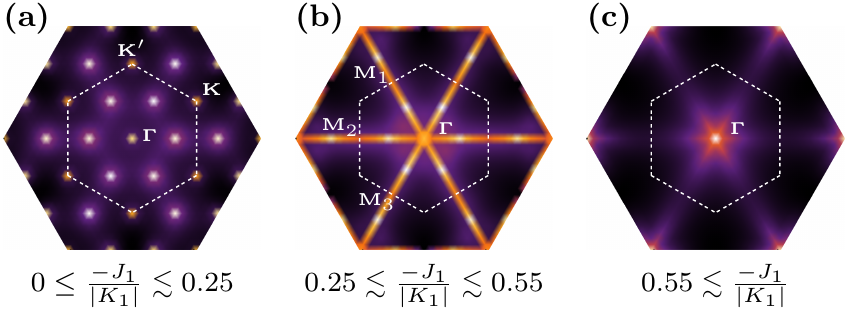}
\caption{Low-$T$ MC static structure factor for $J_1$-$K_1$-$\Gamma_1$ model with $(K_1, \Gamma_1) = (-6.8, +9.5)\,\mathrm{meV}$ (Set 3) and different values of $J_1 \leq 0$ at zero external field. The color scale is logarithmic. Bright spots indicate Bragg peaks. The dotted inner hexagon denotes the first Brillouin zone. The Bragg-peak pattern reveals that the ground state for $J_1 = 0$ is not a zigzag antiferromagnet, but a multi-$\vec Q$ state~(a). This remains true upon inclusion of a small ferromagnetic Heisenberg interaction $J_1 < 0$, until there is, for some finite $J_1$, a direct transition towards an incommensurate phase~(b). Upon further increasing $-J_1$, the ground state becomes the simple ferromagnet~(c).}
\label{fig:sf-set3j1}
\end{figure}

Fig.~\ref{fig:sf-set3j1}(a) shows the static structure factor from low-temperature MC simulations for the Parameter Set 3 at zero external field. The ground state is characterized by six major Bragg peaks at $\vec Q = \pm\frac{2}{3}\mathbf{M}_i$, $i=1,2,3$, as well as three minor peaks $\vec Q = \mathbf{K}, \mathbf{K}', \mathbf{\Gamma}$. We have checked that this remains true also upon increasing the ratio $\Gamma_1 / |K_1|$. By contrast, a zigzag antiferromagnet would lead to only three Bragg peaks at the three $\mathbf{M}$ points of the first Brillouin zone. As a further check, we have explicitly verified that the multi-$\vec Q$ state as found in the MC simulations has indeed a lower energy as the zigzag state when both are cooled down to $T \to 0$, as suggested by the low-$T$ structure factor. Consequently, the classical ground state of the $K_1$-$\Gamma_1$ model with $K_1 < 0$ and $\Gamma_1 > 0$ is not a zigzag state, but a multi-$\vec Q$ state. This was missed in the previous analyses.\cite{rau2014, ran2017, wang2016, footnote5}

In order to examine whether the zigzag ground state can be recovered within the nearest-neighbor model with ferromagnetic $K_1 < 0$, we first slightly modify Parameter Set 3 by supplementing it with a ferromagnetic Heisenberg interaction $J_1 < 0$. This way, however, we find at some finite $J_1$ a direct transition towards an incommensurate phase, which on our finite-size system is signaled by bright lines along $\mathbf{\Gamma}$-$\mathbf{M}_i$ in the structure factor as displayed in Fig.~\ref{fig:sf-set3j1}(b). Eventually, at larger values of $-J_1$, the ground state becomes the simple Heisenberg ferromagnet, indicated by the single Bragg peak at $\vec Q = \mathbf \Gamma$ as displayed in Fig.~\ref{fig:sf-set3j1}(c). There is therefore no zigzag ground state for this particular ratio of $\Gamma_1/|K_1|$.

\begin{figure}[tbp]
\includegraphics[width=\linewidth]{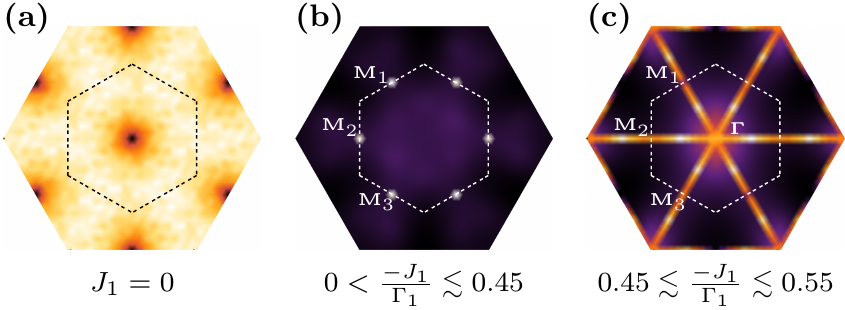}
\caption{Low-$T$ MC static structure factor for $J_1$-$\Gamma_1$ model. For $J_1 = 0$, the ground state is a classical spin liquid,\cite{rousochatzakis2017a} characterized by a structure factor that is finite everywhere except at a few isolated points in the Brillouin zone~(a). By including a small ferromagnetic Heisenberg interaction $J_1 < 0$, the ground-state degeneracy is lifted in favor of the zigzag antiferromagnet, evidenced by its characteristic Bragg peaks at the three $\mathbf{M}$ points~(b). Upon increasing $-J_1$, we recover the incommensurate phase (c), before the ground state is again the ferromagnet at large values of $-J_1$ (not shown).}
\label{fig:sf-j1-gamma1}
\end{figure}

The zigzag ground state can, however, be recovered, without assuming an antiferromagnetic $K_1$ or longer-range interactions (such as $J_3$), by considering a larger ratio $\Gamma_1/|K_1|$ together with a finite ferromagnetic Heisenberg interaction $J_1 < 0$. Fig.~\ref{fig:sf-j1-gamma1} shows the structure factor in the extreme limit $\Gamma_1/|K_1|  \gg 1$ for different values of $J_1 \leq 0$. For $J_1 = 0$ the ground state of this now ``$\Gamma_1$-only model" is a classical spin liquid,\cite{rousochatzakis2017a} characterized by a broad continuum in the structure factor, see Fig.~\ref{fig:sf-j1-gamma1}(a). A small finite $J_1 < 0$ lifts the classical ground-state degeneracy in favor of the zigzag antiferromagnet, indicated by its  characteristic three Bragg peaks at the three $\mathbf M$ points of the first Brillouin zone, see Fig.~\ref{fig:sf-j1-gamma1}(b). At some finite value of $J_1<0$, we again find a transition towards the incommensurate phase, Fig.~\ref{fig:sf-j1-gamma1}(c), before eventually the ferromagnetic ground state is recovered at large values of $-J_1$. There are therefore two distinct mechanisms to stabilize a zigzag ground state in the nearest-neighbor model: Either by strong antiferromagnetic Kitaev interaction $K_1$ supplemented by $J_1 < 0$ and only weak $|\Gamma_1| \ll K_1$ (Scenario 1), or by a strong positive off-diagonal interaction $\Gamma_1 \gg |K_1|$ supplemented by a finite $J_1 < 0$ (Scenario 3).

%%%%%%%%%%%%%%%%%%%%%%%%%%%%%%%%%%%%%%%%%%%%%%%%%%%%%%%%%%%%%%%%%%%%
%%%%%%%%%%%%%%%%%%%%%%%%%%%%%%%%%%%%%%%%%%%%%%%%%%%%%%%%%%%%%%%%%%%%
%%%%%%%%%%%%%%%%%%%%%%%%%%%%%%%%%%%%%%%%%%%%%%%%%%%%%%%%%%%%%%%%%%%%

\end{document}